%% file: root.tex
\documentclass[journal,twoside,web]{ieeecolor}

\usepackage{optidef}

\usepackage{tikzpeople}
\usepackage{generic}
\usepackage{cite}
\usepackage{amsmath,amssymb,amsfonts}
\usepackage{algorithmic}
\usepackage{graphicx}
\usepackage{textcomp}
\usepackage{lipsum,adjustbox}

\usepackage{amssymb,amsmath,color,subfigure}
\usepackage{graphicx}
\usepackage{xspace,stackrel,hyperref}
\usepackage{wrapfig}
\usepackage{mathtools}
\usepackage{tikz,pgfplots}
\usepackage{etoolbox}
\usepackage{autonum}
\usepackage{dsfont}
\usepackage{etoolbox}

\usepackage{multirow}
\usepackage{cite}

\DeclareMathOperator*{\argmin}{arg\,min}

\definecolor{mycolor4}{RGB}{230,97,1}
\definecolor{mycolor2}{RGB}{178,171,210}
\definecolor{mycolor3}{RGB}{253,184,99}
\definecolor{mycolor1}{RGB}{94,60,153}

\makeatletter 
\pretocmd\@bibitem{\color{black}\csname keycolor#1\endcsname}{}{\fail}
\newcommand\citecolor[1]{\@namedef{keycolor#1}{\color{blue}}}
\makeatother

\DeclareMathAlphabet{\mathcal}{OMS}{cmsy}{m}{n}

\def\beq{\begin{equation}}
\def\eeq{\end{equation}}

\newcommand{\mc}{\mathcal}

\newcommand{\Z}{\mathbb{Z}}

\newcommand{\R}{\mathds{R}}
\newcommand{\N}{\mathbb{N}}

\newcommand{\defineas}{\coloneqq}

\newcommand{\norm}[1]{\left\lVert#1\right\rVert}

\definecolor{mycolor1}{RGB}{230,97,1}
\definecolor{mycolor2}{RGB}{178,171,210}
\definecolor{mycolor3}{RGB}{253,184,99}
\definecolor{mycolor4}{RGB}{94,60,153}
\definecolor{mycolor5}{rgb}{0,0,0}

\tikzset{
  pics/car/.style args={#1}{
     code={
     \begin{scope}[scale=0.15]
      \shade[top color=#1, bottom color=white, shading angle={135}]
        [draw=black,fill=red!20,rounded corners=0.2ex] (1.5,.5) -- ++(0,1) -- ++(1,0.3) --  ++(3,0) -- ++(1,0) -- ++(0,-1.3) -- (1.5,.5) -- cycle;
    \draw[ rounded corners=0.5ex,fill=black!20!blue!20!white]  (2.5,1.8) -- ++(1,0.7) -- ++(1.6,0) -- ++(0.6,-0.7) -- (2.5,1.8);
    \draw[thick]  (4.2,1.8) -- (4.2,2.5);
    \draw[draw=black,fill=gray!50,thick] (2.75,.5) circle (.5);
    \draw[draw=black,fill=gray!50,thick] (5.5,.5) circle (.5);
    \end{scope}
     }
  }
}

\newtheorem{assumption}{Assumption}

\newtheorem{theorem}{Theorem}
\newtheorem{proposition}{Proposition}

\newtheorem{definition}{Definition}

\newtheorem{remark}{Remark}

\pgfplotsset{compat=1.17}
\usetikzlibrary{pgfplots.groupplots}

\def\BibTeX{{\rm B\kern-.05em{\sc i\kern-.025em b}\kern-.08em
    T\kern-.1667em\lower.7ex\hbox{E}\kern-.125emX}}

\begin{document}

\title{Hierarchical Pricing Game for Balancing the Charging of Ride-Hailing Electric Fleets}

\author{Marko Maljkovic, Gustav Nilsson, and Nikolas Geroliminis
\thanks{M.~Maljkovic, G.~Nilsson, and N.~Geroliminis are with the School of Architecture, Civil and Environmental Engineering, École Polytechnique Fédérale de Lausanne (EPFL), 1015 Lausanne, Switzerland. {\tt\small \{marko.maljkovic, gustav.nilsson, nikolas.geroliminis\}@epfl.ch}.}
\thanks{Part of
the results of this article appeared in a preliminary form in~\cite{9838005}.}
\thanks{This work was supported by the Swiss National Science Foundation under NCCR Automation, grant agreement 51NF40\_180545. }
}

\maketitle

\begin{abstract}
Due to the ever-increasing popularity of ride-hailing services and the indisputable shift towards alternative fuel vehicles, the intersection of the ride-hailing market and smart electric mobility provides an opportunity to trade different services to achieve societal optimum. In this work, we present a hierarchical, game-based, control mechanism for balancing the simultaneous charging of multiple ride-hailing fleets. The mechanism takes into account sometimes conflicting interests of the ride-hailing drivers, the ride-hailing company management, and the external agents such as power-providing companies or city governments that will play a significant role in charging management in the future. The upper-level control considers charging price incentives and models the interactions between the external agents and ride-hailing companies as a Reverse Stackelberg game with a single leader and multiple followers. The lower-level control motivates the revenue-maximizing drivers to follow the company operator's requests through surge pricing and models the interactions as a single leader, multiple followers Stackelberg game. We provide a pricing mechanism that ensures the existence of a unique Nash equilibrium of the upper-level game that minimizes the external agent's objective at the same time. We provide theoretical and experimental robustness analysis of the upper-level control with respect to parameters whose values depend on sensitive information that might not be entirely accessible to the external agent. For the lower-level algorithm, we combine the Nash equilibrium of the upper-level game with a quadratic mixed integer optimization problem to find the optimal surge prices. Finally, we illustrate the performance of the control mechanism in a case study based on real taxi data from the city of Shenzhen in China. 
\end{abstract}


\section{Introduction}
\label{sec:introduction}
\IEEEPARstart{T}{he} recent increase in affordability of alternative fuel vehicles, and their prospective positive long-term effects on the environment, have paved the way for a steep incline in the number of electric vehicles (EVs) on the streets \cite{iea}. In parallel, the growing popularity of the various services offered by ride-hailing companies justifies the importance of their existence among the transportation services offered within a city. As a result, it is likely that EVs will soon constitute the central part of the fleets operated by ride-hailing companies. Inevitably, the charging management on a company level will become an important factor in maintaining a successful business in the ride-hailing market. 

The availability of the charging infrastructure and the average charging duration remain some of the main reasons preventing EVs' overall dominance over gas-powered vehicles \cite{9318522}. In order to profitably operate large electric fleets, it is likely that the ride-hailing companies would have to devise intelligent coordinated charging strategies subject to the constraints imposed by the temporal distribution of the demand and the power grid supply. In a more distant future, where the company would operate autonomous vehicles, the operator would have the complete flexibility to minimize the company's operational costs. However, in the current society where the driver's daily profit is directly proportional to the passenger kilometers travelled, it is clear that some monetary-based incentives would have to be proposed in order to motivate the drivers to follow the operator's desires. The ride-hailing companies already offer their drivers access to different service facilities, so it is not unlikely that they will offer discounted charging dictated by the subsidies provided by the central authorities, e.g., the government or the power-providing companies. On one hand, through charging incentives the company can maximize the availability of the services by motivating the drivers to charge before the demand peaks. On the other hand, the asymmetric spatial distribution of request origins and destinations makes certain parts of the region less attractive for the drivers. Through properly designed pricing incentives, the central authority could try to increase the coverage and help fight congestion due to a large number of circulating vehicles without passengers \cite{DiaoArticle,BEOJONE2021102890}. 

Inspired by this vision, we present a demand-based, bi-level, pricing system for load balancing the vehicles of different ride-hailing fleets (such as Uber, Lyft, etc.) among public charging stations. We study a scenario in which a central authority has a clear preference about how the vehicles should spread out among the stations, i.e., it defines the set points that describe a desirable EV distribution in an attempt to either help fight congestion or to balance the demand on the grid. 

On the upper level, the central authority is incentivizing the companies to follow the set points through pricing while the company operators try to minimize their operational costs, which among others, include the costs of charging. Due to the limited capacity of the shared charging infrastructure, it is in every company's best interest to decrease the queuing time at the stations. Since the charging stations in the demand-attractive areas are more likely to become overcrowded, there is an inherent competition between the companies, establishing a fertile ground for game theoretic analysis. 

\begin {figure*}
\centering
\begin{adjustbox}{max height=0.45\textwidth, max width=\textwidth}

    \begin{tikzpicture}[scale=0.8]

    \node[draw, shape=circle](ch1) at (7.5,2.5){$\mc M_{1}$};
    \node[draw, shape=circle](ch2) at (7.5,0){$\mc M_{2}$};
    \node[draw, shape=circle](ch3) at (7.5,-2.5){$\mc M_{3}$};
    
       \node (veh1) at (4.5,0.5) {\begin{tikzpicture}\draw (0,0) pic{car=mycolor1}; \end{tikzpicture}};
       \node (veh2) at (4.5,2.5) {\begin{tikzpicture}
 \draw (0,0) pic{car=mycolor1}; \end{tikzpicture}};
       \node (veh3) at (4.5,4.5) {\begin{tikzpicture}
 \draw (0,0) pic{car=mycolor1}; \end{tikzpicture}};

        \node (d11) at (4.7, 1.5)[circle,fill,inner sep=0.75pt]{};
        \node (d12) at (4.7, 1.2)[circle,fill,inner sep=0.75pt]{};  \node (d13) at (4.7, 1.8)[circle,fill,inner sep=0.75pt]{};

       \node (veh4) at (4.5,-0.5) {\begin{tikzpicture}
 \draw (0,0) pic{car=mycolor2}; \end{tikzpicture}};
       \node (veh5) at (4.5,-2.5) {\begin{tikzpicture}
 \draw (0,0) pic{car=mycolor2}; \end{tikzpicture}};
       \node (veh6) at (4.5,-4.5) {\begin{tikzpicture}
 \draw (0,0) pic{car=mycolor2}; \end{tikzpicture}};

        \node (d21) at (4.7, -3.5)[circle,fill,inner sep=0.75pt]{};
        \node (d22) at (4.7, -3.2)[circle,fill,inner sep=0.75pt]{};  \node (d23) at (4.7, -3.8)[circle,fill,inner sep=0.75pt]{};      
        \shadedraw[top color= mycolor1, bottom color=white, draw=mycolor1] (-4.7, 3.2) rectangle (2.2, 1.8);

        \shadedraw[top color= mycolor2, bottom color=white, draw=mycolor2] (-4.7, -1.8) rectangle (2.2, -3.2);
        
        \draw[] (0.5, 3) rectangle (2,2);
        \node (c1m) at (1.25, 2.5){$K_{M}^{1}$};
 
        \draw[] (0.5, -3) rectangle (2,-2);
        \node (c2m) at (1.25, -2.5){$K_{M}^{2}$};

        \draw[] (-4.5, 3) rectangle (-3,2);
        \node (c1f) at (-3.75, 2.5){$K_{F}^{1}$};
 
        \draw[] (-4.5, -3) rectangle (-3,-2);
        \node (c2f) at (-3.75, -2.5){$K_{F}^{2}$};
        
        \draw (-10.5, 0.75) rectangle (-7, -0.75);
        \node[text width=4cm] (ca) at (-7.75, 0){Central authority};
        
        \draw (ch1.east) -- (9, 2.5);
        \draw (ch2.east) -- (9, 0);
        \draw (ch3.east) -- (9, -2.5);
        \draw (9, 2.5) -- (9, -2.5);
        
        \draw (9, 2.5) -- (9, 5.5) -- (-3.75, 5.5);
        \draw (9, -2.5) -- (9, -5.5) -- (-3.75, -5.5);
        \draw[->] (-3.75, 5.5) -- (-3.75, 3);
        \draw[->] (-3.75, -5.5) -- (-3.75, -3);
        \node[mycolor1] (param1) at (-2.25, 6){$A_1,B_1,c_1,D_1,f_1$};
        \node[mycolor2] (param2) at (-2.25, -6){$A_2,B_2,c_2,D_2,f_2$};
        
        \draw[dashed] (0.3, 4.9) rectangle (5.3,0.1);
        \draw[dashed] (0.3, -0.1) rectangle (5.3,-4.9);
        
        \draw[dashed] (-10.7, 4.9) rectangle (-2.8, -4.9);
        
        \draw[->] (-7, 0) -- (-4.5, 2.5);
        \draw[->] (-7, 0) -- (-4.5, -2.5);
        \draw[<->] (-3.75, 2) -- (-3.75, -2);
        
        \node[mycolor1,rotate=45] (price1) at (-5.95, 1.7){$p_1(x^1,x^2)$};
        \node[mycolor2,rotate=-45] (price2) at (-5.95, -1.7){$p_2(x^1,x^2)$};
        
        \draw[->] (-3,2.5) -- (0.5,2.5);
        \draw[->] (-3,-2.5) -- (0.5,-2.5);
        
        \draw[->](2,2.5) -- (4,0.5);
        \draw[->](2,2.5) -- (4,2.5);
        \draw[->](2,2.5) -- (4,4.5);

        \draw[->](2,-2.5) -- (4,-0.5);
        \draw[->](2,-2.5) -- (4,-2.5);
        \draw[->](2,-2.5) -- (4,-4.5);
        
        \draw[->] (5.2, 4.5) -- (ch1.west);
        \draw[->] (5.2, 2.5) -- (ch2.west);
        \draw[->] (5.2, 0.5) -- (ch1.west);
        
        \draw[->] (5.2, -4.5) -- (ch3.west);
        \draw[->] (5.2, -2.5) -- (ch3.west);
        \draw[->] (5.2, -0.5) -- (ch1.west);        
        
        \node[] (p11) at (2.7, 3.8){$\rho^{1}_1$};
        \node[] (p12) at (3, 2.8){$\rho^{2}_1$};
        \node[] (p13) at (2.7, 1.1){$\rho^{\left|\mc V_1\right|}_1$};

        \node[] (p21) at (2.7, -3.9){$\rho^{\left|\mc V_2\right|}_2$};
        \node[] (p22) at (3, -2.2){$\rho^{2}_2$};
        \node[] (p23) at (2.7, -1.2){$\rho^{1}_2$};
        
        \node[] (n1) at (-1.25, 2.8){$x^{1*}$};
        \node[] (n1) at (-1.25, -2.2){$x^{2*}$};
        
        \node[] (G11) at (1.25, 4.5){Game $\mc G_2$};
        \node[] (G12) at (1.25, -0.5){Game $\mc G_2$};
        \node[] (G0) at (-9.75, 4.5){Game $\mc G_1$};
        
        \node[businessman, shirt=mycolor1](b1) at (-2.4, 3.9){};
        \node[businessman, shirt=mycolor2](b2) at (-2.4, -1.1){};
        
        \node[mycolor1, text width=2.5cm](cop1) at (-0.3, 3.8){Operator Company 1};
        \node[mycolor2, text width=2.5cm](cop2) at (-0.3, -1.2){Operator Company 2};        
{\tiny }
\end{tikzpicture}
\end{adjustbox}
    \caption{Schematic sketch of the problem setting with companies $\mc C=\left\{\mc C_1,\mc C_2\right\}$ and charging stations $\mc M=\left\{\mc M_1, \mc M_2, \mc M_3\right\}$. The central body, e.g., the government or the power company, wants to balance the vehicle load on different charging stations by properly setting the pricing policies $p_i(x^1,x^2):\mc X \rightarrow \R^3$ for $i\in\left\{1,2\right\}$. Under the provided pricing policies, each ride-hailing company wants to minimize its own operational cost by steering its set of vehicles $\mc V_i$ to different charging stations. The blocks $K_F^i$  calculate the optimal splits $x^{i*}\in\mc X_i\subseteq\mc P_{\mc M}$ of the ride-hailing fleets in a decentralized manner with little information exchange and based on the parameters $A_i$, $B_i$, $c_i$, $D_i$, $f_i$. The blocks $K_M^i$ calculate the surge price vectors $\rho^v_i\in\R_+^{\left|\mc M\right|}$ for every vehicle $v\in\mc V_i$.} 
\label{fig:problem}
\end{figure*}
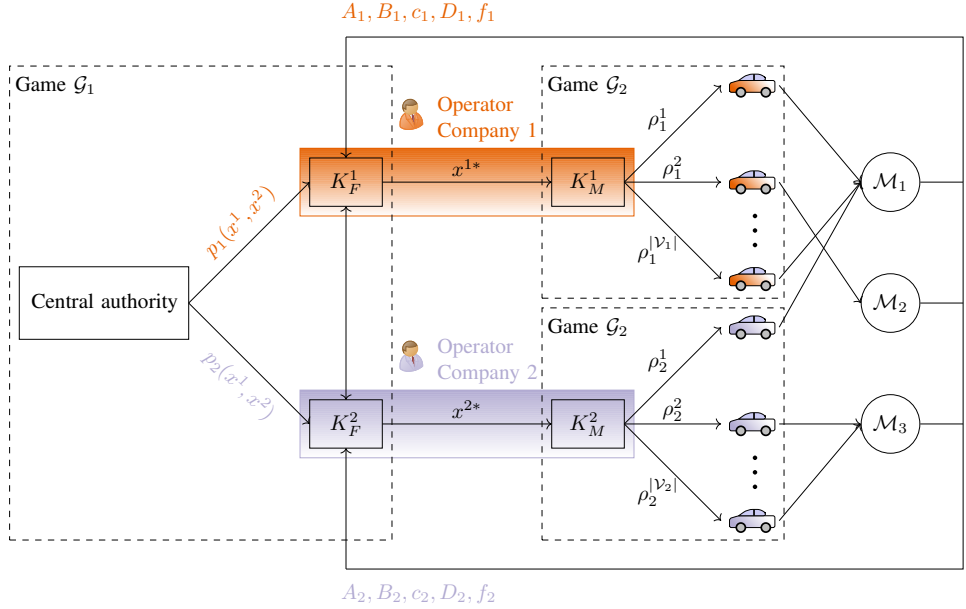

On the lower level, the company's objective is to convince rational drivers to choose a particular charging station in order to match the vehicle distribution dictated by the upper-level algorithm. To do so, the company operator adjusts the ride fares so that the drivers' expected profit is maximized by choosing the stations in accordance with the output of the upper-level module. A complete schematic representation of the problem is presented in Figure~\ref{fig:problem}.      

Various aspects of this problem have been separately studied in the literature. Extensive research in the domain of game theoretic control systems elucidates the effectiveness of such models in solving the problems within the realm of smart mobility. Pricing mechanisms are usually studied in the context of maximizing revenue. Papers \cite{9102356, article4, article5} model the charging stations as selfish revenue-maximizing agents, whereas \cite{9564501} describes a game for modeling the regulatory aspects such as taxes and transportation prices as well as the operational matters for mobility service providers. From the perspective of congestion control in urban networks, different tolling mechanisms, congestion taxes, and congestion-aware routing schemes have been proposed in \cite{article1, article2, 9029550, 9599370} based on the congestion game vehicle routing. Even though the upper-level goal of our method could be identically chosen, our underlying pricing model lies at the intersection of aggregative games and the inherent leader-follower structure present in the Stackelberg and reverse Stackelberg games \cite{Leitmann1978OnGS}. The aggregative nature of the competition between the companies is similar to the one used in \cite{Paccagnan2016a, Paccagnan2019, EVsCharging, 9599370} for charging control of a population of EVs. The Stackelberg game setups in \cite{article4, article5} propose using fixed prices for charging the users, whereas in our system we utilize a feedback pricing method based on the decision of the companies. On par with the Stackelberg-based approaches, there is literature that proposes mechanisms based on reverse Stackelberg games for solving the hierarchical control, bi-level toll design, or coordinated electricity purchase from the power grid for multiple EVs \cite{6823141, article7, 9079179}. From the methodological point of view, these works are similar to ours as they design a decision function for the leader rather than a decision value as in the conventional Stackelberg games. This kind of approach allows us to directly influence the placement of the Nash equilibrium through pricing incentives. However, it also implies that the companies do not know the prices before deciding how to direct their vehicles but rather how their joint decision will influence the charging cost.  

This paper is a continuation of the work presented in \cite{9838005} where the addressed upper-level problem was first introduced. We adopt the game theoretic coordination of the ride-hailing fleets under the pricing game as the upper-level controller and use the computed Nash equilibrium as the set point for the individual EVs-to-station matching controllers. We base these controllers on the concept of surge pricing, which in general refers to an opportunistic adjustment of the price of a service directly depending on the demand for it. In the field of transportation, it is mainly used by ride-hailing companies for maximizing revenue with respect to the temporal and spatial distribution of the demand \cite{7817882, 8258070, 8948099}. In this paper, the operator adjusts the ride fares for each vehicle such that when drivers choose a charging station in order to maximize their profit, the resulting vehicle distribution matches the one dictated by the upper-level controller. 

To the best of our knowledge, the literature does not provide
a comprehensive end-to-end solution for coordinated charging of EV fleets operated by ride-hailing
companies so as to achieve the objective of a central authority. The main contributions of this paper can be summarized as follows:
\begin{itemize}
    \item We analyze the ride-hailing market from the joint perspective of three hierarchical levels of agents and design a decentralized bi-level game theoretic method that yields a no regret solution for the central authority, the companies, and the drivers under the reachability constraints imposed by the battery state of the vehicles.
    \item In order to achieve its optimum, the central authority needs to receive some information about the cumulative state of the EV fleets. We provide a detailed robustness analysis of the proposed solution when only an estimate of the parameters is available to the central authority.
    \item The complete bi-level control system is tested on a new case study developed based on real taxi data from the city of Shenzhen. 
\end{itemize}

The paper is outlined as follows: the rest of this section is devoted to introducing some basic notation. In Section~\ref{sec:model} we introduce the model and state the main formulation. In the following section, Section~\ref{sec:pricing}, we revise the pricing mechanism introduced in \cite{9838005} along with the decentralized algorithm used to compute the Nash equilibrium and perform a detailed robustness analysis. We present the matching module based on surge pricing in Section~\ref{sec:matching} and demonstrate in Section~\ref{sec:example} the performance of the end-to-end bi-level model on a case study based on real taxi data. Finally, Section~\ref{sec:conclusion} concludes the paper  with some ideas for future research.

\subsection{Notation}\label{subsec:not}

Let $\R$ denote the set of real numbers, $\R_+$ the set of non-negative reals and $\Z_+$ the set of non-negative integers. Let $\mathbf{0}_{m}$ and $\mathbf{1}_{m}$ denote the all zero and all one vectors of length $m$ respectively, and $\mathbb{I}_{m}$ the identity matrix of size $m \times m$. For a finite set $\mc A$, we let $\R_{(+)}^{\mc A}$ and $\Z_{(+)}^{\mc A}$ denote the sets of (non-negative) real and integer vectors indexed by the elements of $\mc A$, $\left|\mc A\right|$ the cardinality of $\mc A$ and we let $\mc P_{\mc A}$ be the probability space over the set, i.e., $\mc P_{\mc A} \defineas \{ x \in \R_+^{\mc A} \mid \sum_{i \in \mc A} x_i = 1\}$. Furthermore, for finite sets $\mc A$, $\mc B$ and a set of $|\mc B|$ vectors $x^i\in \R_{(+)}^{\mc A}$, we define $x \defineas \left[x^{i}\right]_{i\in \mc B}\in \R^{|\mc A||\mc B|}$ to be their concatenation and $x^{-i} \defineas \left[x^{j}\right]_{j\in \mc B \setminus i}$ to be the concatenation of all vectors except the one indexed with $i$. For $A\in \R^{n\times n}$, $A\succ 0 (\succeq 0)$ on set $\mc X$ is equivalent to $x^TAx>0 (\geq 0)$ for all $x\in\mc X$. For a diagonal matrix $A \in \R^{n \times n}$, we let $A^*$ denote its pseudo-inverse, i.e., 
\begin{equation}
A^*_{ii} \defineas 
\begin{cases} 1/A_{ii} & \textrm{if } A_{ii} \neq 0, \\
0 & \textrm{otherwise,}
\end{cases}\quad 1 \leq i \leq n\,.
\end{equation}

\section{System model}\label{sec:model}
The problem setup assumes that a set of ride-hailing companies operates in the same region and has access to common charging stations for their EVs. We let $\mc C$ be the set of all ride-hailing companies, and $N_i >0$ the number of vehicles belonging to each company $i \in \mc C$ that need to be charged. The vector $N\in\R^{\mc C}_+$ stores the number of EVs of every company and $m_c=\left|\mc C\right|$ is the number of ride-hailing companies. We denote the set of all charging stations as $\mc M$ and their total number as $m_s=\left|\mc M\right|$. We define $M_j>0$ as the number of available spots for simultaneous charging at the station $j\in\mc  M$, i.e., the charging station's capacity and denote the vector of all charging stations capacities as $M \in \R_+^{\mc M}$. 

For every company $i \in \mc C$, we let $\mc V_i$ be the set of its vehicles with $\left|\mc V_i\right|=N_i$. To describe how the company operator wants to distribute the vehicles among charging stations, i.e., the company's decision variable, we use a continuous allocation vector $x^{i} \in \mc P_{\mc M}$. Here, $x^i_j$ is the fraction of vehicles from company $i \in \mc C$ that will be sent to charging station $j \in \mc M$. In addition, we define for each company a discrete allocation vector $n^i\in\Z_{+}^{\mc M}$, with $n^{i}_{j}\in\Z_{+}$ describing the actual integer number of vehicles that the operator of the fleet would send to station $j$ based on the value of $x^{i}_{j}$. Since some drivers cannot reach some of the charging stations due to their battery level and hence not all choices of $x^{i}$ are feasible, we define for each company the feasibility sets 
\begin{equation}
    \mc F_{j}^{i} \defineas \left\{v \in \mc V_i \mid v \text{ can reach station } j \right\} \,.
    \label{eq:fsets}
\end{equation}
For each $i\in\mc C$,
we need to define how the company operator can choose the discrete allocation vector $n^i$ in order to achieve a many-to-one matching between the vehicles $\mc V_i$ and the charging stations $\mc M$. Hence, a proper set of admissible continuous allocation vectors $x^i$ also has to be defined. In general, the operator should always be able to assign either $n^i_j=\left\lfloor{N_{i}x^{i}_{j}}\right\rfloor$ or $n^i_j=\left\lceil{N_{i}x^{i}_{j}}\right\rceil$ vehicles to station $j\in\mc M$ in an attempt to satisfy the global constraint $\mathbf{1}_{m_s}^Tn^i=N_i$, without the concern that the existence of the perfect matching between the vehicles and the charging stations will be violated. We summarize this in Definition~\ref{def:admisdisx} where we formally introduce an admissible discrete allocation vector $n^i$. 
\medskip
\begin{definition}\label{def:admisdisx}
    For each $i\in\mc C$, a discrete allocation vector $n^i\in\Z^{\mc M}_+$ is admissible to the operator if 
    \begin{itemize}
        \item for every $j\in\mc M$, $n^i_j=\left\lfloor{N_{i}x^{i}_{j}}\right\rfloor$ or $n^i_j=\left\lceil{N_{i}x^{i}_{j}}\right\rceil$ ,
        \item $\mathbf{1}_{m_s}^Tn^i=N_i$,
        \item it is possible to assign each vehicle $v\in\mc V_i$ to exactly one charging station in $\mc M$ in order to respect the discrete vehicle distribution $n^i$.
    \end{itemize}
\end{definition}
\medskip
In Section~\ref{sec:pricing}, we will analytically construct for every $i\in\mc C$ a set $\mc X_{i}\subseteq \mc P_{\mc M}$ of admissible continuous allocation vectors $x^i$ that will respect the feasibility sets $\mc F^i_j$ and guarantee existence of at least one admissible $n^i$. With the existence of those sets, we proceed to define $\mc X\defineas \prod_{i\in\mc C} \mc X_{i}$ and $\mc X_{-i}\defineas \prod_{j\in\mc C\setminus i} \mc X_{j}$. 

We stack the decision vectors of all companies into $x \defineas \left[x^{i}\right]_{i\in \mc C}\in \mc X$ and define $x^{-i} \defineas \left[x^{j}\right]_{j\in \mc C \setminus i}\in\mc X_{-i}$ as the stacked decision vectors of all companies except the company~$i$. The distribution of all vehicles among charging stations is described by the vector $\sigma\left(x\right) \defineas \sum_{i \in \mc C} N_{i}x^{i} \in \R_+^{\mc M}$, i.e., $\sigma\left(x\right)_j$ denotes the total number of vehicles that are directed to station $j\in\mc M$. In addition, we let $\sigma\left(x^{-i}\right) \defineas \sum_{j \in \mc C \setminus i} N_j x^j \in \R_+^{\mc M}$ describe the distribution of all vehicles except for those belonging to company $i$. Finally, for every vehicle $v\in\mc V_i$, we let $x^{v}\in\left\{0,1\right\}^{\mc M}$ denote a one-hot encoding of driver's charging station choice and $\Omega_{v}\subset\left\{0,1\right\}^{\mc M}$ be the set of all one-hot encodings that correspond to vehicle's feasible charging stations. Let us define sets $\Omega_i\defineas\prod_{v\in\mc V_i}\Omega_v$ for every $i\in\mc C$. Then the vector $\mu^i\defineas \left[x^v\right]_{v\in\mc V_i}\in\Omega_i$ of stacked drivers' decisions describes the perfect matching between the EVs and the charging stations and the vector $\sigma\left(\mu^i\right)\defineas\sum_{v\in\mc V_i}x^v$ describes the integer number of vehicles directed to each charging station.

\subsection{Central authority's objective}

The central authority is ranked the highest in the agent hierarchy and, as explained in the introductory section, is interested in balancing the vehicles so as to optimize a system-level objective, e.g., to try to help fight the congestion in the city or to balance the demand on the power grid. In order to ease the notation, throughout the rest of the paper we will refer to the central authority as the government. The model that we consider allows the government to balance the vehicles so as to minimize any system-level objective of the form

\begin{equation}
    J_G (\sigma\left(x\right)) = \frac{1}{2}\sigma\left(x\right)^{T}A_{G} \sigma\left(x\right)+b_{G}^{T}\sigma\left(x\right) \,,
    \label{eq:JG}
\end{equation} for some diagonal weight matrix $A_{G}\succ 0$ and $b_{G}\in \R^{\mc M}$. We analyze a special case of~\eqref{eq:JG} that corresponds to balancing the ride-hailing vehicles to help decrease the congestion induced by the idle vehicles cruising around the charging stations in demand-attractive regions. We assume that based on the historical data about the demand, the government can a priori decide on a desired total vehicle distribution $\mc Z\in\mc P_{\mc M}$ and is therefore interested in balancing the vehicles so that the total number of vehicles directed to each of the charging stations is given by $\hat{N}\in\R^{\mc M}_+$, i.e., to minimize
\begin{equation}
    J_G (\sigma\left(x\right)) = \frac{1}{2}\left\|\sigma(x)-\widehat{N}\right\|^{2}_{2,A_{G}} \,,
    \label{eq:JG2}
\end{equation}
where the set point is given by $\widehat{N}=\mathbf{1}_{m_s}^T N\mc Z$. By choosing $b_{G}=-A_{G}\widehat{N}\in \R^{\mc M}$, we make the problem of minimizing~\eqref{eq:JG2} a special case of~\eqref{eq:JG}.

The government interacts with the ride-hailing companies through pricing policies. In order to steer the companies to the system optimum, the government will assign an individual pricing policy to each company for each charging station. The policies announced to company $i\in\mc C$ are a function of all companies' choices and can be jointly represented by a mapping $p_{i}\left(x^{i},x^{-i}\right) : \mc X_{i}\times \mc X_{-i}\rightarrow \R^{\mc M} $.

\subsection{Company's objective}

Every ride-hailing company is interested in minimizing its operational cost under the feasibility constraints imposed by the battery status of its vehicles. In Figure~\ref{fig:problem}, the individual blocks $K^i_F$ represent the cost minimizing controllers in charge of optimally splitting the ride-hailing fleets. We model the cost as the sum of three terms: the expected queuing cost, the charging cost, and the negative expected revenue. 

The expected queuing cost of company $i\in\mc C$ depends on its personal choice but also on the vehicle distribution of all other companies, i.e., on $x^i$ and $\sigma\left(x^{-i}\right)$. We adopt the expected queuing cost model
\begin{equation}
    \label{eq:J1true}
    \begin{split}
        J_{1}^{i}\left(x^{i}, \sigma(x^{-i})\right)&=N_{i}\left(x^{i}\right)^{T}Q\left(N_ix^i+\sigma(x^{-i})-M\right)\\
        &= N_{i}\left(x^{i}\right)^{T}Q\left(\sigma(x)-M\right)\,,
    \end{split}
\end{equation}
where $Q\in\R^{\mc M\times \mc M}$ is a positive definite diagonal scaling matrix whose diagonal entries depict how expensive it is to queue in the regions around charging stations. The charging stations located in the city's more busy areas should experience higher queuing costs and hence have a higher corresponding diagonal entry in the $Q$ matrix. In addition, the more the capacity of the station is exceeded, the higher the cost should be which is directly enabled through the inner product with the vector $\sigma\left(x\right)-M$. Moreover, we can write the queuing cost in a general form given by
\begin{equation}
J_{1}^{i}\left(x^{i}, \sigma(x^{-i})\right)=\frac{1}{2} (x^{i})^T A_{i}x^{i}+(x^{i})^TB_{i}\sigma\left(x^{-i}\right)+c_{i}^{T}x^{i} \,,
\label{eq:J1igen}
\end{equation}
for some diagonal matrices $A_{i} \in \R^{\mc M \times \mc M}$, $B_{i} \in \R^{\mc M \times \mc M}$ and vector $c_{i}\in\R^{\mc M}$. It is clear that in order to cast~\eqref{eq:J1true} in the form of~\eqref{eq:J1igen}, we need to set
\begin{equation}
    A_{i}\defineas2N_{i}^{2}Q,\:\:B_{i}\defineas N_{i}Q,\:\:c_{i}\defineas -N_{i}QM \,.
\label{eq:params}
\end{equation}

The charging cost of company $i\in\mc C$ depends on the charging price policy assigned to it and its personal decision, i.e., $p_{i}\left(x^{i},x^{-i}\right)$ and $x^i$. It can be written as
\begin{equation}
    J_{2}^{i}\left(x^{i}, p_{i}\left(x^{i},x^{-i}\right)\right)=(x^{i})^T D_{i} p_{i}\left(x^{i},x^{-i}\right) \,,
\label{eq:J2}
\end{equation}
where $D_{i} \in \R^{\mc M \times \mc M}$ is diagonal, $D_i\succeq 0$ and the entries $\left(D_{i}\right)_{kk}$ can be interpreted as the part of the total charging demand of the company $i$ to be served at the station $k$. The value $\left(D_{i}\right)_{kk}$ is calculated based on the average charging demand per vehicle, averaged over the set of vehicles that can reach station $k$ with their current battery level and under the chosen battery discharge model.

We let the negative expected revenue encompass the information about the difference between the cost of vehicles being idle while travelling to the charging stations and the expected profit in the regions around the charging stations. We consider this an important aspect of the analysis as by integrating the cost of being idle and the expected profit, we can model the scenarios in which a company might decide to pay more for charging in the regions of higher demand and hence, higher expected profit. We assume the negative expected revenue is a function of only the company's personal choice $x^i$ that can be modelled as a linear cost $J_{3}^{i}\left(x^{i}\right)=f_i^Tx^i$. In Section~\ref{sec:example}, we provide a detailed description of how all the parameters can be calculated for a real-world scenario.

It will be shown in the next section that the system optimal pricing policies $p_{i}\left(x^{i},x^{-i}\right)$ require that the company shares information about the parameters of the cost function. We assume the government provides a large enough fixed subsidy to every company that is willing to help attain the system optimum. Consequently, we assume it is in every company's best interest to disclose true information about the private parameters. Nevertheless, as the parameter $D_i$ provides sensitive information about the charging cost, in Section~\ref{sec:pricing} we provide a detailed robustness analysis when only an estimate of this parameter is available. Because we assume a fixed subsidy, it does not take part in the optimization procedure so the total cost of the company is given by
\begin{multline}
\label{eq:companyloss}
    J^{i}\left(x^{i},x^{-i}, p_{i}\left(x^{i},x^{-i}\right)\right)  = \\ J_{1}^{i}\left(x^{i}, \sigma(x^{-i})\right)
     +J_{2}^{i}\left(x^{i}, p_{i}\left(x^{i},x^{-i}\right)\right)+ J_{3}^{i}\left(x^{i}\right) \,.
\end{multline}

\subsection{Driver's objective}

For any $i\in\mc C$, the drivers $v\in\mc V_i$, are ranked the lowest in the agent hierarchy. After the upper level controllers $K_F^i$ of all $i\in\mc C$ determine the optimal car fleet splits $x^*$, each operator will pick $n^{i}$ and calculate the vector of charging prices $p_i^*\defineas p_i^*\left(x^{i*},x^{-i*}\right)$. In order to achieve the optimal cost imposed by the output of the upper-level controller $K_F^i$, the operator has to motivate the drivers in the fleet to pick charging stations such that $n^{i}$ is attained. To do so, we assume the operator utilizes an additional degree of freedom in its operation management, i.e., the surge pricing. 

Within a particular company $i\in\mc C$, we let the operator pick for every driver $v\in\mc V_i$ a vector $\rho^{v}\in\Omega_{\rho}^v\subseteq\mathbb{R}^{\mc M}_{+}$ such that $\left(\rho^{v}\right)_j\in\mathbb{R}_+$ represents the value of a supplementary fare added to the standard one if the driver chooses to operate in the region around the charging station $j$. The achieved vehicle distribution after matching EVs with charging stations is determined by $\mu^i$ so the lower level controller $K_M^i$, as shown in Figure~\ref{fig:problem}, has to determine the optimal values of $\rho^v $ for all $v\in\mc V_i$ so as to minimize the cost function
\begin{equation}
    J^i_M\left(\sigma\left(\mu^i\right) |\: n^i\right)=\frac{1}{2}\norm{\sigma\left(\mu^i\right)-n^{i}}_{2}^{2}\,.
    \label{eq:Jmi}
\end{equation}
where $J^i_M\left(\cdot \:|\: n^i\right)$ means that the company operator's objective function is parameterized by $n^i$.

On the other hand, we assume each driver chooses a charging station so as to minimize the personal cost that comprises three terms: the charging cost, the negative expected revenue should the driver choose to operate for a predefined time interval $\tau^v\in\R_+$ in the same region where it charged  and the bonus profit due to surge pricing. For any driver $v\in\mc V_i$, the general form of its cost is given by 
\begin{equation}
    J^{v}\left(x^{v},\rho^{v}\right)=\left(x^{v}\right)^TD_vp_i^*+\left(x^{v}\right)^Tg_v - \left(x^{v}\right)^TH_v\rho^{v}\,,
\label{eq:costdriver}
\end{equation}
where $D_v\in\R^{\mc M \times \mc M}$ is diagonal, $D_v\succeq 0$ and $\left(D_v\right)_{kk}$ describes the vehicle's expected charging demand if it chooses the charging station $k$, $g_v\in\R^{\mc M}$ is the original negative expected revenue if operating for a time interval of $\tau^v$ and $H_v\in\R^{\mc M\times\mc M}_+$ describes the expected bonus profit the driver will receive due to surge pricing. A detailed description of all parameters in a realistic scenario is given in Section~\ref{sec:example}. 

Having defined the objectives and the decision variables of all agents in the system, we proceed to formally define the problem in the next section.

\subsection{Problem formulation}

Figure~\ref{fig:problem} shows a detailed schematic representation of different interactions between the agents. Each company $i\in\mc C$ would like to allocate its vehicles according to 
\begin{equation}
    x^{i*} \in \argmin_{x^{i} \in \mc X_i} J^{i}\left(x^{i},x^{-i},p_{i}\left(x^{i},x^{-i}\right)\right) \,. 
    \label{eq:companymin}
\end{equation}
with the individual objective functions defined as in~\eqref{eq:companyloss}. If the pricing policies $p_{i}\left(x^{i},x^{-i}\right)$ are fixed for every $i\in\mc C$, then solving~\eqref{eq:companymin} boils down to  allocating the vehicles according to a game defined by
\begin{equation}
    \mc G_0\defineas \left\{\min_{x^{i}\in \mc X_{i}}J^{i}\left(x^{i},x^{-i}\right),\forall i\in \mc C\right\} \,,
    \label{eq:gameform}
\end{equation}
whose Nash equilibrium $x^*$ is given in Definition~\ref{def:NE}.
\medskip
\begin{definition}[$\varepsilon$-Nash equilibrium]\label{def:NE}
    A joint strategy $x^*\in\mc X$ is an $\varepsilon$-Nash equilibrium of the game $\mc G_0$, if there exists an $\varepsilon>0$ such that for all $i\in\mc C$ and for all $x^i\in\mc X_i$ it holds that
    \begin{equation}
        J^i\left(x^{i*}, x^{-i*}\right)\leq J^i\left(x^{i}, x^{-i*}\right)+ \varepsilon \,.
        \label{eq:NEdef}
    \end{equation}
    If~\eqref{eq:NEdef} holds with $\varepsilon=0$ then $x^*$ is a Nash equilibrium.
\end{definition}
\medskip

We say that the companies admit a system optimum $x^*$ under the pricing policies $p_{i}\left(x^{i},x^{-i}\right)$ if $x^*$ minimizes~\eqref{eq:JG} and is a Nash equilibrium of the game $\mc G_0$ at the same time. Let the concatenation of all the pricing policies be denoted as
\begin{equation}
    \pi\left(x\right)\defineas\left[p_{i}\left(x^{i},x^{-i}\right)\right]_{i\in\mc C}\,,
\end{equation}
and $\Omega_{G}$ be the space of concatenated pricing policies. The problem of finding the system optimal pricing policies that minimize~\eqref{eq:JG} falls under the category of reverse Stackelberg games (RSG) \cite{6402334} with one leader, i.e., the government, and multiple followers, i.e., the companies, introduced in Definition~\ref{def:RSG}.
\medskip
\begin{definition}[Reverse Stackelberg game]\label{def:RSG}
    Let $x^{i*}\in\mc X_i$ and $\pi^*\left(x\right) : \mc X \rightarrow \R^{Nm_s}$. Then a tuple $\left(\pi^*\left(x\right), x^{*}\right)$ is a solution of a reverse Stackelberg game with one leader and multiple followers if it solves a bi-level optimization problem $\mc G_1$:
    \begin{equation}
        \mc G_1:=\left\{\begin{array}{c}
        \pi^*\left(\cdot\right)\in\displaystyle\argmin _{\pi\left(\cdot\right) \in \Omega_{G}}  J_{G}\left(\sigma\left(x^*\right)\right) \\
        \text { s.t. } x^{i*} \in \displaystyle\argmin_{x^{i} \in \mc X_i} J^{i}\left(x^{i},x^{-i},p^*_{i}\left(\cdot\right)\right), \forall i \in\mc C
        \end{array}\right\} \,.
        \label{eq:RSG}
    \end{equation}
\end{definition}
\medskip

The Nash equilibrium of game $\mc G_0$ under the system optimal pricing policies will provide the set points for the lower level controllers $K^i_M$. Let $\Omega^i_\rho\defineas\prod_{v\in\mc V_i}\Omega^v_{\rho}$ and $\rho^i\defineas\left[\rho^v\right]_{v\in\mc V_i}\in\Omega^i_{\rho}$. Then for every company $i\in\mc C$, the problem of finding~$\rho^v$ for every $v \in\mc V_i$ corresponds to centralized computation of the leader's optimal strategy in a Stackelberg game with one leader, i.e., the company operator, and multiple followers, i.e., the drivers, formally introduced in Definition~\ref{def:SG}.
\medskip
\begin{definition}[Stackelberg game]\label{def:SG}
    Let $\mu^{i*}\in\Omega_{i}$ and $\rho^{i*}\in\Omega_{\rho}^i$. Then a tuple $\left(\rho^{i*}, \mu^{i*}\right)$ is a solution of a Stackelberg game with one leader and multiple followers if it solves a bi-level optimization problem $\mc G_2$:
    \begin{equation}
        \mc G_2:=\left\{\begin{array}{c}
        \rho^{i*}\in\displaystyle\argmin _{\rho^i \in \Omega_{\rho}^i}  J_{M}^i\left(\sigma\left(\mu^{i*}\right) |\: n^i\right) \\
        \text { s.t. } x^{v*} \in \displaystyle\argmin_{x^{v} \in \Omega_v} J^{v}\left(x^v,\rho^{v*}\right), \forall v \in\mc V_i
        \end{array}\right\} \,.
        \label{eq:SG}
    \end{equation}    
\end{definition}
\medskip

In the next section, we will show that it is possible to construct a pricing mechanism that solves $\mc G_1$ and yields a unique system optimum if the constraint sets $\mc X_i$ are compact and convex. Furthermore, we will show in Section~\ref{sec:matching} that for the Nash equilibrium of $\mc G_0$ induced by the system optimal pricing policies, it is always possible to solve game $\mc G_2$ such that $\left(\rho^{i*}, \mu^{i*}\right)$ is a global minimizer of $J_{M}^i\left(\sigma\left(\mu^{i*}\right)|\:n^i\right)$ defined in~\eqref{eq:Jmi}. In other words, we will show that it is always possible to find surge prices $\rho^{i*}$ for all $i\in\mc C$ such that the resulting EV distribution described by $\mu^{i*}$ perfectly matches the desired one, i.e., $\sigma\left(\mu^{i*}\right)=n^i$.

\section{Upper-level control -- pricing mechanism}\label{sec:pricing}

At the beginning of this section, we propose the admissible sets $\mc X_i$ for each $i\in\mc C$. They will
guarantee that for any $x^i\in\mc X_i$, the operator will have at least one admissible discrete allocation $n^i\in\Z^{\mc M}_+$ corresponding to Definition~\ref{def:admisdisx}. Moreover, we will later show that these admissible sets allow us to construct pricing policies yielding a unique system optimum. 

\subsection{Admissible continuous allocation vectors}\label{subsec:admis}

We start by analyzing the conditions guaranteeing that each company operator $i\in\mc C$ will be able to perfectly match each vehicle $v\in\mc V_i$ with a charging station in $\mc M$. Assuming that each vehicle can reach at least one charging station with the current battery level, the following theorem provides a necessary and sufficient condition on the discrete allocation vector $n^i$ that guarantees the existence of a perfect matching. 
\medskip
\begin{theorem}\label{th:match}
    For each company $i\in\mc C$, let feasibility sets $\mc F^i_j$ be defined as in~\eqref{eq:fsets} and $n^i\in\Z^{\mc M}_+$ denote the discrete allocation vector. There exists a many-to-one matching between the vehicles $v\in\mc V_i$ and the charging stations $\mc M$ if and only if for every subset $\mc S \subseteq \mc M$ it holds that 
    \begin{equation}
    \sum_{j\in \mc S}n_{j}^{i}\leq \left|\bigcup_{j \in \mc S} \mc F_{j}^{i}\right|\,.
    \label{eq:disccond}
    \end{equation}
\end{theorem}
\medskip
\begin{proof}
We look at a bipartite graph $G_{i}=\left(\mc V_{i}\cup \mc S_{i} , E_{i}\right)$ where~$\mc S_{i}$ is defined as $\mc S_{i}=\bigcup_{j\in \mc M}\mc S_{i}^{j}$ such that for all $j_{1},j_{2}\in \mc M$, $j_{1}\neq j_{2}$ it holds that $\mc S_{i}^{j_{1}}\cap \mc S_{i}^{j_{2}}=\emptyset$. Each $\mc S_{i}^{j}$ is comprised of~$n_{j}^{i}$ copies of the vertex that corresponds to the charging station~$j$. The set of edges $E_{i}$ is formed such that $v\in \mc V_{i}$ is connected to  $s\in \mc S_{i}^{j}$ if $v\in \mc F_{j}^{i}$. The two sets have equal number of vertices $\left|\mc V_{i}\right|=N_{i}=\sum_{j\in\mc M}n^{i}_{j}=\left|\mc S_{i}\right|$ which means that desired matching is possible if and only if there exists an $\mc S_{i}$-perfect matching on graph $G_{i}$. Since condition \eqref{eq:disccond} corresponds exactly to the condition of Hall's marriage theorem~\cite{hall}, the equivalence is proved.
\end{proof}
\medskip
The inequality~\eqref{eq:disccond} is intuitive as it states that for any subset of the charging stations, the company operator must not allocate more vehicles than what is available. With this in mind, in the following proposition we show how to analytically construct sets $\mc X_i\subseteq\mc P_{\mc M}$.
\medskip
\begin{proposition}\label{prop:Kbar}
For each company $i \in \mc C$, define the set $\mc X_{i}\subseteq \mc P_{\mc M}$ such that $x^{i}\in\mc X_{i}$ if for all proper subsets $\mc S$ of $\mathcal{M}$, it holds that
\begin{equation}
 N_{i} \sum_{j \in\mc S} x^{i}_{j} \leq \max \left\{0,\left|\bigcup_{j \in\mc S} \mc F_{j}^{i}\right|-|\mc S|\right\}  \,.
 \label{eq:setxi}
\end{equation}If the state of the car fleet does not correspond to a degenerate case for which $\mc X_{i}=\emptyset$, then every $x^{i}\in\mc X_{i}$ yields an admissible $n^i$ corresponding to Definition~\ref{def:admisdisx}. 
\end{proposition}
\medskip
\begin{proof}
 We now show that if $x^{i}$ satisfies~\eqref{eq:setxi} then any $n^{i}$ in accordance with Definition~\ref{def:admisdisx} satisfies the assumption given by \eqref{eq:disccond}. We distinguish 2 cases: $\mc S\subset \mc M$ and $\mc S=\mc M$. For $\mc S\subset \mc M$ we can write $$\sum_{j\in\mc  S}n_{j}^{i}=\sum_{j\in P_{1}}\left\lfloor{N_{i}x^{i}_{j}}\right\rfloor + \sum_{j\in P_{2}}\left\lceil{N_{i}x^{i}_{j}}\right\rceil$$
where $P_{1}\cup P_{2}=\mc S\wedge P_{1}\cap P_{2}=\emptyset$. We have
$\sum_{j\in P_{1}}\left\lfloor{N_{i}x^{i}_{j}}\right\rfloor\leq \sum_{j\in P_{1}}N_{i}x^{i}_{j}$ and
$\sum_{j\in P_{2}}\left\lceil{N_{i}x^{i}_{j}}\right\rceil= \sum_{j\in P_{2}}N_{i}x^{i}_{j}+\left\{N_{i}x_{j}^{i}\right\}$
where $\forall j \in P_{2}$ it holds that $\left\{N_{i}x_{j}^{i}\right\}\leq 1$. We have $$\sum_{j\in \mc S}n_{j}^{i}\leq \sum_{j\in P_{1}\cup P_{2}}N_{i}x^{i}_{j}+\sum_{j\in P_{2}}\left\{N_{i}x_{j}^{i}\right\}\leq \sum_{j\in \mc S}N_{i}x^{i}_{j}+\left|P_{2}\right|$$ which combined with \eqref{eq:setxi} finally gives
$$\sum_{j\in\mc S}n_{j}^{i}\leq \left|\bigcup_{j \in\mc  S} \mc F_{j}^{i}\right| - \left|\mc S\right|+\left|P_{2}\right|\leq \left|\bigcup_{j \in\mc  S}\mc F_{j}^{i}\right|$$ because $\left|P_{2}\right|\leq \left|\mc S\right|$. For $\mc S=\mc M$ we have that the condition given by \eqref{eq:disccond} is fulfilled with the equality since $\sum_{j\in\mc  S}n_{j}^{i} = N_{i}=\left|\bigcup_{j \in\mc  S}\mc F_{j}^{i}\right|$. The case when for some $\mc S$ it holds that $\left|\bigcup_{j \in\mc  S} \mc F_{j}^{i}\right| - \left|\mc S\right|\leq 0$ leads to $x_{j}^{i}=0$ for all $j\in\mc  S$, which in return leads to $n_{j}^{i}=0$, so no matching is required. 
\end{proof}
\medskip
In general, the degenerate states of the ride-hailing fleet manifest that the drivers have very limited options when choosing which charging station to take due to their current battery level. These cases are not of interest to us so from this point on we assume $\mc X_i\neq\emptyset$ for every $i\in\mc C$. This means that for each company, the set of admissible continuous allocations defined as in Proposition~\ref{prop:Kbar}, represents the intersection of a probability space and $2^{m}-2$ linear inequalities given by \eqref{eq:setxi} making it compact and convex. As a result, for the constraint sets~$\mc X_i$ as in Proposition~\ref{prop:Kbar}, the pricing policies in the following subsection will give rise to a unique system optimum. 

\subsection{System optimal pricing policies}\label{subsec:policy}

This section introduces the pricing mechanism that will solve the reverse Stackleberg game $\mc G_1$ between the government and the ride-hailing companies given in Definition~\ref{def:RSG}. We will show that for the admissible sets $\mc X_i$ defined as in Proposition~\ref{prop:Kbar} and under the proposed pricing mechanism, the game $\mc G_0$ introduced in Definition~\ref{def:NE}, will have a unique Nash equilibrium that simultaneously minimizes the government cost given by~\eqref{eq:JG}. Hence, solving for the Nash equilibrium of the game between the companies will directly correspond to minimizing the government's objective explaining why we refer to the proposed pricing policies as the system optimal. 
\medskip
\begin{definition}[System Optimal Pricing Policies]\label{def:pricingpolicies}
For each company $i \in \mc C$, let $D_{i}^{*}$ be the pseudo-inverse of $D_i$ and  
\begin{equation}
    p_{i}\left(x^{i},x^{-i}\right)=D_{i}^{*}\left[\frac{1}{2}\overline{\text{A}}_{i} x^{i}+ \overline{\text{B}}_{i}\sigma\left(x^{-i}\right)+\Delta_{i}\right] \,,
    \label{eq:optimalpolicy}
\end{equation}
where $\overline{\text{A}}_{i}=N_{i}^{2} A_{G}-A_{i}$ , $\overline{\text{B}}_{i}=N_{i} A_{G}-B_{i}$ and $\Delta_{i}= N_{i}b_{G}-c_{i}-f_{i}$.
\end{definition}
\medskip
For every company $i \in \mc C$, it could be the case that some stations are unreachable for every vehicle $v\in\mc V_i$. In such a scenario, we let the corresponding diagonal entry in the matrix $D_i$ be equal to zero, making the matrix non-invertible. Conversely, since the company $i$ will not use these charging stations, letting the prices for them be equal to zero through the pseudo-inverse defined in Section~\ref{subsec:not} will not affect the solution of the problem. 

In the following theorem, we show that the system optimal pricing policies adhere to a unique Nash equilibrium of the game played between the companies.
\medskip
\begin{theorem}
For all companies $i \in \mc C$, let the sets $\mc X_i$ be designed as in Proposition~\ref{prop:Kbar}. Then, with the system optimal pricing policies in Definition~\ref{def:pricingpolicies}, the game $\mc G_0$ in~\eqref{eq:gameform} has a unique Nash equilibrium. 
\label{th:t2}
\end{theorem}
\medskip
\begin{proof}
We start by inserting policy \eqref{eq:optimalpolicy} into \eqref{eq:companyloss} and observing that for $x^{i}\in\mc X_{i}$ it holds that $D_{i}D_{i}^{*}x^{i}=x^{i}$. This  transforms the cost of each company $i\in\mc C$ into:
\begin{equation}
    \label{eq:modcost}
    \begin{split}
    J^{i} (x^i, x^{-i}) &=\frac{1}{2}\left(x^{i}\right)^{T}N_{i}^{2}A_{G}x^{i}+ \\
    &+\left(x^{i}\right)^{T}N_{i}A_{G}\sigma\left(x^{-i}\right)+ \\
    &+\left(x^{i}\right)^{T}N_{i}b_{G} \,.
    \end{split}
\end{equation}
It is evident that the cost function~\eqref{eq:modcost} is continuous in $x\in\mc X$, but because $A_GN_i\succ 0$ by design,  it is also quadratic and convex in $x^i\in\mc X_i$ for any fixed $x^{-i}\in\mc X_{-i}$. For non-degenerate states of the car fleets, the action spaces $\mc X_i$ are compact, convex, and satisfy Slater's constraint by construction, hence, the game admits a Nash equilibrium~\cite[T.1]{Rosen}. Let us define 
\begin{equation}
    g(x,r)\defineas \left[-r_{i}\nabla_{x_{i}}J^{i}(x)\right]_{i\in \mc C}\,,
    \label{eq:gxr}
\end{equation}
where $x\in\mc X$ and $r=\left[r_{i}\right]_{i\in\mc C}\in \R^{\mc C}_{>0}$. A sufficient condition for the uniqueness of the Nash equilibrium \cite[T.2]{Rosen} is that the matrix $\Gamma\in\R^{m_c\times m_c}$, $\Gamma\defineas G(x,r)+G^{T}(x,r)$
be negative definite for all $x\in\mc X$ and some $r\in \R^{\mc C}_{>0}$, with $G(x,r)$ being the Jacobian of $g(x,r)$ with respect to $x$. For $r=\textbf{1}_{m_c}$ and any $x\in \mc X$ we have
\begin{equation}
    x^{T}\Gamma x=-2\left(\sum_{i\in \mc C}N_{i}x^{i}\right)^{T}A_{G}\left(\sum_{i\in \mc C}N_{i}x^{i}\right)\,.
\end{equation}
Since $\mc X_{i}\subseteq\mc P_{\mc M}$ for all $i \in \mc C$, we have that $\sum_{i\in \mc C}N_{i}x^{i}\neq \textbf{0}_{|\mc M|}$. Because $A_{G}\succ 0$, we have  $x^{T}\Gamma x<0$ for all $ x\in\mc X$ which proves that $\Gamma$ is negative definite on $\mc X$ and that the Nash equilibrium is unique.
\end{proof}
\medskip
\begin{remark}\label{rem:r1}
    For $r=-\textbf{1}_{m_c}$, equation~\eqref{eq:gxr} represents the pseudo-gradient of the game $\mc G_0$, i.e., $F\left(x\right)\defineas \left[\nabla_{x_{i}}J^{i}(x)\right]_{i\in \mc C}$. If we define $A^{T}\defineas\left[N_{i}\mathbb{I}_{m_s}\right]_{i\in \mc C}\in\R^{m_sm_c\times m_s}$, then the pseudo-gradient can be written as
    \begin{equation}
    F(x)=A^{T}A_{G}A x+A^{T}b_{G} \,,
    \label{eq:gamemap}
\end{equation}
which is affine in $x$, i.e., $F\left(x\right)=F_1x+F_2$ for $F_1\defineas A^{T}A_{G}A$ and $F_2\defineas A^{T}b_{G}$. Based on the properties of $F_1$, we will later present a decentralized computation scheme that is sure to converge to a Nash equilibrium of the game. 
\end{remark}
\medskip
In the following theorem, we proceed to show that the unique Nash equilibrium of the game $\mc G_0$ will minimize the government objective given by~\eqref{eq:JG}. 
\medskip
\begin{theorem}\label{t:t3}
For all companies $i \in \mc C$, let the sets $\mc X_i$ be designed as in Proposition~\ref{prop:Kbar}. Then, with the system optimal pricing policies in Definition~\ref{def:pricingpolicies}, the Nash equilibrium $x^{*}$ of $\mc G_0$ defines a tuple $(\pi\left(x\right), x^{*})$ that solves the game $\mc G_1$, i.e.,  
\begin{equation}
    x^{*} \in \argmin_{x \in \mc X} J_{G}\left(\sigma(x)\right) \,. 
\end{equation}
\end{theorem}
\medskip
\begin{proof}
For the matrix $A$ defined in Remark~\ref{rem:r1}, the government optimization problem is equivalent to 
\begin{equation}
    \min_{x\in \mc X}\:J_{G}(x)\defineas\frac{1}{2}x^{T}A^{T}A_{G}Ax+b_{G}^{T}Ax \,.
    \label{eq:JGx}
\end{equation}
The Hessian of the government loss is given by $$\nabla^{2}_{x}J_{G}(x)=A^{T}A_{G}A\,.$$ Since for all $x^{i} \in \mc X_{i}\subseteq\mc P_{\mc M}$ we have $\sum_{i\in \mc C}N_{i}x^{i}\neq \textbf{0}_{m_s}$ and $$x^{T}A^{T}A_{G}Ax=\left(\sum_{i\in\mc C}N_{i}x^{i}\right)^{T}A_{G}\left(\sum_{i\in\mc C}N_{i}x^{i}\right)> 0\,,$$ the government loss function is convex since $\nabla^{2}_{x}J_{G}(x)\succeq 0$. According to \cite[4.21]{ConvexOptimization}, the Nash equilibrium $x^{*}$ of the game $\mc G_0$ will be the minimizer of \eqref{eq:JGx} on $\mc X$ if and only if
\begin{equation}
    \left\langle\nabla_{x} J_{G}(x)\mid_{x=x^{*}}, y-x^{*}\right\rangle \geq 0,\:\forall y \in \mc X \,.
    \label{eq:conditionmin}
\end{equation}
Under the pricing policies defined in \eqref{eq:optimalpolicy}, $J_{G}(x)$ is the exact potential \cite{PotentialGames} for game $\mc G_0$, satisfying for any fixed $ x^{-i}\in\mc X_{-i}$
\begin{equation}
    \nabla_{x^{i}}J^{i}\left(x^{i},x^{-i}\right)=\nabla_{x^{i}}J_{G}\left(x^{i},x^{-i}\right) \,, \:\: \forall x^{i}\in \mc X_{i} \,.
    \label{eq:potential}
\end{equation}
Inserting~\eqref{eq:potential} into~\eqref{eq:conditionmin} transforms the condition into
\begin{equation}
    \sum_{i\in\mc C}\left\langle\nabla_{x^{i}} J^i(x)\mid_{x=x^*}, y^{i}-x^{i*}\right\rangle\geq 0,\:\forall y^i \in \mc X_i \,.
    \label{eq:condt}
\end{equation}
Because $x^*$ is the Nash equilibrium, we have that for all $ i\in\mc C$ it holds that $x^{i*}\in\argmin_{x^{i}\in\mc X_{i}}J^{i}\left(x^{i},x^{-i*}\right)$. Consequently, we have that $\left\langle\nabla_{x^{i}} J^{i}(x)\mid_{x=x^*}, y^{i}-x^{i*}\right\rangle \geq 0$ holds for all $ y^{i}\in\mc X_{i}$ and all $i\in\mc C$ according to~\cite{ConvexOptimization}, which ensures that the optimality condition~\eqref{eq:condt} is satisfied.
\end{proof}
\subsection{Decentralized computation of the Nash equilibrium}
The decisions spaces $\mc X_i$ reflect the current state of the car fleets, i.e., their battery status and world position. In reality, these sets are in general private, hence not known to the government, which calls for a decentralized algorithm for computation of the Nash equilibrium. Algorithms that require little information exchange between the agents have been analyzed in \cite{Decentralized}. Based on the properties of the pseudo-gradient of the game $\mc G_0$, different, distributed, iterative schemes can converge to a Nash equilibrium of an aggregative game. Because the game map  in~\eqref{eq:gamemap} is not strictly monotonic since $F_1=A^TA_GA\succeq 0$, the commonly used Picard-Banach iteration procedure used for finding a fixed point of a mapping will not guarantee convergence to a Nash equilibrium. Instead, we use an algorithm based on the  Krasnoselskij iteration that guarantees convergence for non-expansive mappings~\cite{ApproxFixPoint}. The procedure is described in the following proposition.
\medskip
\begin{proposition}\label{prop:gamma}
Under the system optimal pricing policies and for sets $\mc X_i$ as in Proposition~\ref{prop:Kbar}, for every $\gamma$ such that
\begin{equation}
    0<\gamma<\frac{2}{\lambda_{\text{max}}\left(F_1\right)} \,,
    \label{eq:optgamma}
\end{equation}
a distributed iterative scheme given by
\begin{equation}
    x^{i}_{k+1}=\frac{1}{2}\left(x^{i}_{k}+\Pi_{\mc X_{i}}\left[x^{i}_{k}-\gamma \nabla_{x^{i}} J^{i}\left(x^{i}_{k}, x^{-i}_{k}\right)\right]\right) \,, 
\label{eq:iterativeproc}
\end{equation}
where $\Pi_{\mc X_{i}}$ denotes the projection operator onto $\mc X_{i}$, converges to the Nash equilibrium of the game $\mc G_0$. 
\end{proposition}
\medskip
\begin{proof}
According to the theory of Variational Inequalities~\cite{VISurvey}, a point $x^*\in\mc X$ is a Nash equilibrium of game $\mc G_0$ with the game map $F(x)$ defined by \eqref{eq:gamemap} if and only if
\begin{equation}
    F\left(x^*\right)^{T}\left(y-x^*\right)\geq 0\,,
    \label{eq:cond1}
\end{equation}
holds for all  $y \in\mc X$. Condition~\eqref{eq:cond1} is equivalent to 
\begin{equation}
    x^*=\Pi_{\mc X}\left[x^*-\gamma F\left(x^*\right)\right]\,.
    \label{eq:cond2}
\end{equation}
Indeed,~\eqref{eq:cond2} is equivalent to
\begin{equation}
    x^*\in\argmin_{z\in\mc X}\norm{z-\left(x^*-\gamma F(x^*)\right)}^{2}_{2}\,.
    \label{eq:cond3}
\end{equation}
According to~\cite{VISurvey}, since $\mc X$ is convex,~\eqref{eq:cond3} is equivalent to 
\begin{equation}
    2\left(z-\left(x^*-\gamma F(x^*)\right)\right)^{T}(y-z)\geq 0\,, \quad \forall y \in\mc X\,.
\end{equation}
 It is clear that setting $z=x^*$ in equation~\eqref{eq:cond3} shows that $x^*$ is a Nash equilibrium of game $\mc G_0$ if and only if it is a fixed point of the mapping $H(x)=\Pi_{\mc X}\left[x-\gamma F\left(x\right)\right]$. For the Krasnoselskij iteration $x_{k+1}=0.5\left(x_{k}+H\left(x_{k}\right)\right)$ to converge, $H(x)$ has to be non-expansive and $x_0\in\mc X$. The projection operator is non-expansive so for $H(x)$ to be non-expansive, it suffices to choose $\gamma$ such that $\bar{H}(x)=\mathbb{I}x-\gamma F(x)=\left(\mathbb{I}-\gamma F_{1}\right)x-\gamma F_{2}$ is non-expansive. Since $\bar{H}(x)$ is affine, it suffices to choose $\gamma$ such that $\norm{\mathbb{I}-\gamma F_{1}}_{2}\leq1$. Because $F_{1}$ is symmetric, this is equivalent to
 \begin{equation}
     \max_{i}\left|\lambda_{i}\left(\mathbb{I}-\gamma F_{1}\right)\right|\leq1\,.
     \label{eq:cond4}
 \end{equation}
Since $F_{1}\succeq 0$, then for $\gamma$ given in \eqref{eq:optgamma} the condition~\eqref{eq:cond4} is satisfied because $-1\leq 1-\gamma\lambda_{i}\left( F_{1}\right)\leq 1$ holds for all eigenvalues of $F_1$. Therefore, $\bar{H}(x)$ is non-expansive and thus $H(x)$ is non-expansive too. 
\end{proof}
\subsection{Robustness analysis}\label{subsec:robust}
As introduced in the previous section, the system optimal pricing mechanism depends on the willingness of the company operators to share information about the personal cost functions. The parameters $A_i$, $B_i$, and $c_i$ are inherently known to the government as they are either a government's personal choice or they characterize the charging infrastructure and the region in which the ride-hailing companies operate. On the other hand, the parameters $D_i$ and $f_i$ describe the average state of the company's fleet, i.e., the average battery level and the average position of the vehicles, and as such are in general not at the government's disposal. In Section~\ref{sec:example}, we will explain in detail how $f_i$ can be calculated if the position of the ride-hailing vehicles is known. The parameter $D_i$ represents an even more sensitive piece of information. Apart from the position of the vehicles, calculating the charging demand through matrix $D_i$ necessitates providing information about the current battery level of the vehicles in the fleet. In this section, we are interested in analyzing what are the consequences if the government has only an estimate of the parameter $D_i$.

For every $i\in\mc C$, let $D_i^{\Delta}\in\R^{\mc M\times\mc M}$ be a diagonal matrix that describes the discrepancy between the true value of the parameter $D_i$ and the estimated value known to the government. In the following definition, we introduce the approximate system optimal pricing policies.
\medskip
\begin{definition}[Approximate Optimal Pricing Policies]\label{def:apppricingpolicies}
For each company $i \in \mc C$, let $D_i^{\Delta}\in\R^{\mc M\times\mc M}$ and  
\begin{equation}
    \widetilde{p}_{i}\left(x^{i},x^{-i}\right)=\left(D_{i}^{*}+D_i^{\Delta}\right)\left[\frac{1}{2}\overline{\text{A}}_{i} x^{i}+ \overline{\text{B}}_{i}\sigma\left(x^{-i}\right)+\Delta_{i}\right] \,,
    \label{eq:pertpolicy}
\end{equation}
where $\overline{\text{A}}_{i}=N_{i}^{2} A_{G}-A_{i}$ , $\overline{\text{B}}_{i}=N_{i} A_{G}-B_{i}$ and $\Delta_{i}= N_{i}b_{G}-c_{i}-f_{i}$.
\end{definition}
\medskip
The cost function of each company $i\in\mc C$ under the approximate optimal pricing policies is now given by:
\begin{equation}
    \widetilde{J}^i\left(x^i,x^{-i}\right)=J^i\left(x^i,x^{-i}\right)+\left(x^i\right)^TD_i^2D_i^{\Delta}p_i\left(x^i,x^{-i}\right)\,.
    \label{eq:approxcost}
\end{equation}
The derivative of the company's cost function under the approximate optimal policies with respect to the personal decision variable is given by
\begin{equation}
    \nabla_{x^{i}}\widetilde{J}^{i}\left(\cdot\right)=\nabla_{x^{i}}J_{i}\left(\cdot\right) + D_iD_i^{\Delta}\left[\overline{\text{A}}_{i} x^{i}+ \overline{\text{B}}_{i}\sigma\left(x^{-i}\right)+\Delta_{i}\right]\,.
\end{equation}
Hence, the pseudo-gradient of the game $\mc G_0$ under the approximate  optimal prices can be written as
\begin{equation}\label{eq:tildef}
    \widetilde{F}\left(x\right)=F\left(x\right)+\Phi\left(L_1x+L_2\right)=F\left(x\right)+\Delta F\left(x\right)\,,
\end{equation}
where $L_1$, $L_2$ and $\Phi$ are given by:
\begin{equation}
    L_{1}\defineas\left[\begin{array}{cccc}D_{1} \overline{\text{A}}_{1} & \overline{\text{B}}_{1} N_{2} & \cdots & \overline{\text{B}}_{1} N_{m_c} \\ \overline{\text{B}}_{2} N_{1} & D_{2} \overline{\text{A}}_{2} & \cdots & \overline{\text{B}}_{2} N_{m_c} \\ \vdots & \vdots & \ddots & \vdots \\ \overline{\text{B}}_{m_c} N_{1} & \cdots & \overline{\text{B}}_{m_c} N_{m_c-1} & D_{m_c} \overline{\text{A}}_{m_c}\end{array}\right]\,,
\end{equation}
$\Phi\defineas\text{blockdiag}\left(\{D_i^{\Delta}\}_{i\in \mc C}\right)$ and $L_2\defineas\left[D_i\Delta_i\right]_{i\in \mc C}$. In Theorem~\ref{th:t2}, we exploited the convexity of the company's loss function to prove the existence of the Nash equilibrium. With approximate optimal pricing policies as in Definition~\ref{def:apppricingpolicies}, the existence is not guaranteed as $D_i^{\Delta}$ alters the quadratic term of the cost given in~\eqref{eq:approxcost}. Hence, for the robustness analysis we adopt the following assumption.
\medskip
\begin{assumption}\label{ass:1}
For every company $i\in\mc C$, it holds that 
\begin{equation}
    N_i^2\left(\mathbb{I}_{m_c}+D_iD_i^{\Delta}\right)A_G-D_iD_i^{\Delta}A_i\succeq0\,.
    \label{eq:ass}
\end{equation}
\end{assumption}
\medskip
With an analogous line of thought as in Theorem~\ref{th:t2}, Assumption~\ref{ass:1} guarantees the cost functions given by~\eqref{eq:approxcost} are convex in $x^i$. Hence, the existence of a Nash equilibrium of game $\mc G_0$ with the approximate optimal pricing policies as in Definition~\ref{def:apppricingpolicies} is also guaranteed.
\medskip

We can now show that the Nash equilibrium of the game $\mc G_0$ under the approximate optimal pricing is an $\varepsilon$-Nash equilibrium of $\mc G_0$ under the system optimal pricing policies. Let
\begin{equation}
    t_x^i\defineas\left[N_i\left(x^i\right)^T,\sigma\left(x^{-i}\right)^T\right]^T\in\mc T_i\,.
\end{equation}
Since $\mc X_i$ is compact and convex, then $N_ix^i$ also belongs to a compact and convex set. Moreover, since the Minkovski sum preserves convexity, $\sigma\left(x^{-i}\right)$ also belongs to a compact and convex set making $\mc T_i$ compact and convex as well. The cost function $J^i\left(x^i,x^{-i}\right)$ is quadratic in $t_x^i$ 
\begin{equation}
    J^i\left(x^i,x^{-i}\right)=\frac{1}{2}\left(t_x^i\right)^T\overline{A}_Gt_x^i+\left(t_x^i\right)^T\overline{b}_G\,,
\end{equation}
where $\overline{A}_G\defineas\left[\begin{array}{cc}A_G & A_G \\ A_G & \mathbf{0}_{m_s\times m_s}\end{array}\right]$ and $\overline{b}_G\defineas\left[b_G^T,\mathbf{0}_{m_s}^T\right]^T$. Hence, $J^i\left(t_x^i\right)\defineas J^i\left(x^i,x^{-i}\right)$ is $\eta_i$-Lipschitz continuous over the set $\mc T_i$ for some $\eta_i\in\R_+$, i.e., 
\begin{equation}
    \norm{J^i\left(t_x^i\right)-J^i\left(t_{\overline{x}}^i\right)}_2\leq\eta_i\norm{t_x^i-t_{\overline{x}}^i}_2\,,
\end{equation}
for any $t_x^i,t_{\overline{x}}^i\in\mc T_i$. If we denote $\overline{\eta}\defineas \displaystyle\max_i \eta_i$, then the upper bound on $\varepsilon$ is provided in the following proposition.
\medskip
\begin{proposition}
    For all companies $i\in\mc C$, let the sets $\mc X_i$ be designed as in Proposition~\ref{prop:Kbar}. Then, with the approximate optimal pricing policies in Definition~\ref{def:apppricingpolicies}, the Nash equilibrium $\widetilde{x}$ of $\mc G_0$ is an $\varepsilon$-Nash equilibrium of $\mc G_0$ with the system optimal pricing policies in Definition~\ref{def:pricingpolicies} with \begin{equation}
        \varepsilon\leq 4\overline{\eta}\left(\sum_{i=1}^{N}N_i-\frac{1}{2}\min_i N_i\right)\,.
    \end{equation}
\end{proposition}
\medskip
\begin{proof}
Let us define $\Delta J^i\left(\widetilde{x},x^i\right)\defineas J^i\left(\widetilde{x}^i,\widetilde{x}^{-i}\right)-J^i\left(x^i,\widetilde{x}^{-i}\right)$. According to Definition~\ref{def:NE}, $\widetilde{x}$ is an $\varepsilon$-Nash equilibrium of $\mc G_0$ if for every $i\in\mc C$, $\Delta J^i\left(\widetilde{x},x^i\right)\leq\varepsilon$ holds for every $x^i\in\mc X_i$. Let $x^*\in\mc X$ be the Nash equilibrium of $\mc G_0$ under the system optimal pricing policies. We can now write:
\begin{equation}
    \begin{split}
        \Delta J^i\left(\widetilde{x},x^i\right)&=J^i\left(x^i,x^{-i*}\right)-J^i\left(x^i,\widetilde{x}^{-i}\right) + \\
        & + J^i\left(\widetilde{x}^i,\widetilde{x}^{-i}\right)-J^i\left(x^{i*},x^{-i*}\right) + \\
        & + J^i\left(x^{i*},x^{-i*}\right)-J^i\left(x^i,x^{-i*}\right) \,.
    \end{split}
\end{equation}
Because $x^*\in\mc X$ is the Nash equilibrium of $\mc G_0$, we have that
\begin{equation}
    J^i\left(x^{i*},x^{-i*}\right)-J^i\left(x^i,x^{-i*}\right)\leq 0\,.
\end{equation}
Therefore, we can write
\begin{equation}
    \label{eq:cond5}
    \begin{split}
        \Delta J^i\left(\widetilde{x},x^i\right)&\leq \norm{J^i\left(x^i,x^{-i*}\right)-J^i\left(x^i,\widetilde{x}^{-i}\right)}_2 + \\
        & + \norm{J^i\left(\widetilde{x}^i,\widetilde{x}^{-i}\right)-J^i\left(x^{i*},x^{-i*}\right)}_2\,.
    \end{split}
\end{equation}
Because $x^i_j\geq 0$ for all $j\in\mc M$, it holds that $\norm{N_ix^i}_2\leq\norm{N_ix^i}_1=N_i$, $\norm{\sigma\left(x^{-i}\right)}_2\leq\norm{\sigma\left(x^{-i}\right)}_1\leq\sum_{j\neq i}N_j$ and $\norm{t_x^i}_2\leq\norm{t_x^i}_1=\sum_{i=1}^{j}N_i$. Combining with~\eqref{eq:cond5}, we get
\begin{equation}
    \begin{split}
        \Delta J^i\left(\widetilde{x},x^i\right)&\leq\eta_i\norm{\sigma\left(x^{-i*}\right)-\sigma\left(\widetilde{x}^{-i}\right)}_2+
        \eta_i\norm{t_{\widetilde{x}}-t_{x^*}}_2\leq\\
        &\leq 2\overline{\eta}\sum_{j\neq i}N_j+2\overline{\eta}\sum_{i=1}^{N}N_i=\\
        & =4\overline{\eta}\sum_{i=1}^{N}N_i-2\overline{\eta} N_i\,,
    \end{split}
\end{equation}
which completes the proof.
\end{proof}
\medskip

From the government perspective, it is important to quantify the gap between the values of the government loss function in the Nash equilibrium when the system optimal and approximate optimal pricing mechanisms are applied. Let us assume the following Krasnoselskij iteration procedure is applied
\begin{equation}
    x^{i}_{k+1}=\frac{1}{2}\left(x^{i}_{k}+\Pi_{\mc X_{i}}\left[x^{i}_{k}-\widetilde{\gamma} \nabla_{x^{i}} \widetilde{J}^{i}\left(x^{i}_{k}, x^{-i}_{k}\right)\right]\right) \,.
\label{eq:iterativeprocapprox}
\end{equation}
Analogously to Proposition~\ref{prop:gamma}, for~\eqref{eq:iterativeprocapprox} to converge to a Nash equilibrium, it suffices to choose $\widetilde{\gamma}$ such that
\begin{equation}
   0<\widetilde{\gamma}<\frac{2}{\lambda_{\text{max}}\left(F_1+\Phi L_1\right)}\,.
   \label{eq:gammaprox}
\end{equation}
Let $\overline{K}\in\N$ denote the number of executed updated steps according to iterative scheme given in~\eqref{eq:iterativeprocapprox} with $\widetilde{\gamma}$ given in~\eqref{eq:gammaprox}. The best attained government loss with the iterative procedure based on~\eqref{eq:iterativeprocapprox} and~\eqref{eq:gammaprox} is the one closest to the value of the government loss function in the Nash equilibrium of $\mc G_0$ with the system optimal pricing policies. We have shown in Theorem~\ref{t:t3} that $x^*$ is the global minimizer of the government's loss, hence for all $k\in\N$ it holds that $J_G\left(\sigma\left(x_k\right)\right)-J_G\left(\sigma\left(x^*\right)\right)\geq0$. Therefore, the best attained government loss with the approximate system optimal pricing policies is defined as $$J_G^{\text{best}}\defineas\min_{k\leq\overline{K}}J_G\left(\sigma\left(x_k\right)\right).$$ We are interested in analyzing the difference $J_G^{\text{best}}-J_G\left(\sigma\left(x^*\right)\right)$, where $x^*\in\mc X$ is the Nash equilibrium of $\mc G_0$ under the system optimal pricing policies. Because $\mc X$ is compact and convex, there exists some $R_x\in\R_+$ such that $\norm{x}_2\leq R_x\leq m_{c}$. Note that the pseudo-gradient map of $\mc G_0$ under the approximate optimal pricing policies satisfies
\begin{equation}\label{eq:ftilder}
    \norm{\widetilde{F}\left(x\right)}_2\leq\norm{F_1+\Phi L_1}_2R_x + \norm{F_2+\Phi L_2}_2=R_{\widetilde{F}}\,.
\end{equation}
If we define a mapping 
\begin{equation}
    \psi\left(x\right)\defineas\Delta F^T\left(x\right)\left(x-x^*\right)\,,
    \label{eq:psi}
\end{equation}
then the main robustness result is stated in the following proposition.
\medskip
\begin{proposition}\label{prop:robust}
    For all companies $i\in\mc C$, let the sets $\mc X_i$ be defined as in Proposition~\ref{prop:Kbar}. Let the Krasnoselskij procedure be defined by~\eqref{eq:iterativeprocapprox} and~\eqref{eq:gammaprox}, $x_0\in\mc X$ be the initial value and $x^*\in\mc X$ denote the Nash equilibrium of $\mc G_0$ with the system optimal pricing policies as in Definition~\ref{def:pricingpolicies}. Then after $\overline{K}\in\N$ iterations of the update procedure it holds that
    \begin{equation}
        J_G^{\text{best}}-J_G\left(\sigma\left(x^*\right)\right)\leq\frac{\norm{x_0-x^*}^2_2}{\widetilde{\gamma}\left(\overline{K}+1\right)}+\frac{\widetilde{\gamma}R_{\widetilde{F}}^2}{2}-\frac{\sum_{k=0}^{\overline{K}}\psi(x_k)}{\widetilde{\gamma}\left(\overline{K}+1\right)}\,.
    \end{equation}
\end{proposition}
\medskip
\begin{proof}
    We start by analysing $\norm{x_{k+1}-x^*}_2^2$ and using the equality $2u^Tv=\norm{u}^2+\norm{v}^{2}-\norm{u-v}^2$. Let the mapping $H\left(x\right)$ be defined as $H\left(x\right)=\Pi_{\mc X}\left[x-\widetilde{\gamma} \widetilde{F}\left(x\right)\right]$. Then we have
    \begin{equation}
        \label{eq:norm1}
        \begin{split}
            \norm{x_{k+1}-x^*}_2^2&=\frac{1}{4}\norm{x_{k}-x^*}_2^2+\frac{1}{4}\norm{H\left(x_k\right)-x^*}_2^2+\\
            &+\frac{1}{2}\left(x_k-x^*\right)^T\left(H\left(x_k\right)-x^*\right)\,.
        \end{split}
    \end{equation}
    Equation~\eqref{eq:norm1} is equivalent to
    \begin{equation}
        \label{eq:norm2}
        \begin{split}
            \norm{x_{k+1}-x^*}_2^2&=\frac{1}{2}\norm{x_{k}-x^*}_2^2+\frac{1}{2}\norm{H\left(x_k\right)-x^*}_2^2-\\
            &-\frac{1}{4}\norm{H\left(x_k\right)-x_k}^2_2\,.\end{split}
    \end{equation}
    Because the set $\mc X$ is compact and convex, for every $x\in\R_{+}^{m_cm_s}$ and every $y\in\mc X$, it holds that
    \begin{equation}\label{eq:norm3}
        \norm{\Pi_{\mc X}\left(x\right)-y}^2_2\leq\norm{x-y}^2_2\,.
    \end{equation}
Combining~\eqref{eq:norm3} with~\eqref{eq:norm2} gives
\begin{equation}\label{eq:norm4}
    \norm{x_{k+1}-x^*}_2^2\leq\frac{1}{2}\norm{x_{k}-x^*}_2^2+\frac{1}{2}\norm{x_k-x^*-\widetilde{\gamma}\widetilde{F}\left(x_k\right)}_2^2\,.
\end{equation}
Expanding the right-hand side gives
\begin{equation}\label{eq:norm4}
\begin{split}
    \norm{x_{k+1}-x^*}_2^2&\leq\norm{x_{k}-x^*}_2^2+\frac{\widetilde{\gamma}^2}{2}\norm{\widetilde{F}\left(x_k\right)}^2_2\\
    &-\widetilde{F}^T\left(x_k\right)\left(x_k-x^*\right)\,.
\end{split}
\end{equation}
Combining~\eqref{eq:potential} with the first-order convexity condition for the government's loss function gives
\begin{equation}\label{eq:norm5}
    J_G\left(\sigma\left(x^*\right)\right)\geq J_G\left(\sigma\left(x_k\right)\right)-F^T\left(x_k\right)\left(x_k-x^*\right)\,.
\end{equation}
Plugging~\eqref{eq:tildef} into~\ref{eq:norm5} yields
\begin{equation}\label{eq:norm6}
\begin{split}
    -\widetilde{F}^T\left(x_k\right)\left(x_k-x^*\right) &\leq J_G\left(\sigma\left(x^*\right)\right)-J_G\left(\sigma\left(x_k\right)\right)-\\
    &-\Delta F^T\left(x_k\right)\left(x_k-x^*\right)
\end{split}
\end{equation}
Applying the inequality~\eqref{eq:norm4} recursively and combining it with~\eqref{eq:norm6} and~\eqref{eq:ftilder}, we get
\begin{equation}\label{eq:norm7}
    \begin{split}
        \norm{x_{k+1}-x^*}_2^2&\leq\norm{x_{0}-x^*}_2^2+\sum_{k=0}^{\overline{K}}\left(\frac{\widetilde{\gamma}^2R_{\widetilde{F}}}{2} -\psi\left(x_k\right)\right)+\\
        & +\widetilde{\gamma}\sum_{k=0}^{\overline{K}}\left(J_G\left(\sigma\left(x^*\right)\right)-J_G\left(\sigma\left(x_k\right)\right)\right)\,.
    \end{split}
\end{equation}
Finally, combining~\eqref{eq:norm7} with $\norm{x_{k+1}-x^*}_2^2\geq0$ and $J_G\left(\sigma\left(x_k\right)\right)-J_G\left(\sigma\left(x^*\right)\right)\geq J_G^{\text{best}}-J_G\left(\sigma\left(x^*\right)\right)$ completes the proof.
\end{proof}
\medskip
\begin{remark}
    Mapping $\psi\left(x\right)$ in~\eqref{eq:psi} is quadratic in $x$ and always attains a finite minimum and maximum value over the compact and convex set $\mc X$. However, the convexity properties of the mapping are determined by the definiteness of the matrix $\Phi L_1$, i.e., $\Phi L_1\succeq0$ yields a convex $\psi\left(x\right)$ whereas $\Phi L_1\prec0$ yields a concave one. 
\end{remark}
\medskip
With this we complete the theoretical robustness analysis of the upper-level control modules $K_F^i$, for all $i\in\mc C$. In Section~\ref{sec:example}, we will illustrate the robustness behaviour of the system optimal pricing policies in a scenario based on the real taxi data from the city of Shenzhen, China. Regardless of the extent to which the upper-level control mechanism succeeds in minimizing the government's objective, the lower-level matching controllers $K_M^i$ will always be just in charge of tracking the reference set by $K_F^i$. In the following section we will show that the matching controllers can always choose the surge prices so as to achieve perfect tracking.   

\section{Lower-level control -- matching mechanism}\label{sec:matching}
In Section~\ref{sec:model}, we touch upon the problem of motivating the drivers to pick the stations such that the vehicle distribution dictated by the $K^i_F$ controllers is attained. We assume that the drivers are rational and interested in maximizing their daily profit, hence we let the lower level controllers $K_M^i$ play the game $\mc G_2$ in Definition~\ref{def:SG}. The modules $K_M^i$ essentially constitute a reference tracking mechanism making $\mc G_2$ parameterized by the output of the upper-level controllers $K^i_F$ and constrained by the feasibility sets of the companies. 

If the $n^i$ value is chosen such that~\eqref{eq:disccond} holds for every $i\in\mc C$, then the unconstrained minimization of $J^i_M\left(\sigma\left(\mu^i\right) |\: n^i\right)$ always yields the optimal value of zero. However, the company operator and the drivers play a game so it is the job of $K_M^i$ block to make sure the same minimum is attained. The solution of the game $\mc G_2$ can be found in a centralized manner by solving a mixed integer program with bilinear constraints
\begin{mini}
{x^{v},\rho^v, v\in\mc V_{i}}{J^i_M\left(\sigma\left(\mu^i\right) |\: n^i\right)}
{}{}
\addConstraint{x^{v}\in\argmin_{\tilde{x}^{v}\in\Omega_{v}}J^{v}\left(\tilde{x}^{v},\rho^v\right), \forall v\in\mc V_{i}}{}
\addConstraint{\rho^v\geq \rho_{\text{min}}, \forall v\in\mc V_{i}}{}
\addConstraint{x^{v}\in\Omega_{v}, \forall v\in\mc V_{i}}{}
\addConstraint{\mu^i=\sum_{v\in\mc V_i}x^v}{}
\label{mini:op1}
\end{mini}
where $\rho_{\text{min}}\in\R_+^{\mc M}$ represents the vector of minimal surge prices per region and for every $v\in\mc V_i$, the condition $x^{v}\in\argmin_{\tilde{x}^{v}\in\Omega_{v}}J^{v}\left(\tilde{x}^{v},\rho^v\right)$ can be expressed as, at most, $m_s$ bilinear inequalities. In the following theorem, we will show that the optimal value of the optimization problem~\eqref{mini:op1} is zero. 
\medskip
\begin{theorem}
    For all companies $i\in\mc C$, let the surge-pricing optimization problem be defined as~\eqref{mini:op1} and let $n^i\in\Z_+^{\mc M}$ satisfy~\eqref{eq:disccond}. If the solution of~\eqref{mini:op1} is given by a tuple $\left(\rho^{i*}, \mu^{i*}\right)$ then it holds that
    \begin{equation}
        \label{eq:zeromatchingcost}
        J^i_M\left(\sigma\left(\mu^{i*}\right) |\: n^i\right)=0\,.
    \end{equation}
\end{theorem}
\medskip
\begin{proof}
    It is clear that $J^i_M\left(\sigma\left(\mu^{i*}\right) |\: n^i\right)\geq 0$ so we will show that we can always find $x^{v},\rho^v$ for all $v\in\mc V_{i}$ such that~\eqref{eq:zeromatchingcost} holds. For any $n^i$ chosen under the condition~\eqref{eq:disccond} there exists a perfect matching. This means that there exists a set of $N_i$ vectors $x^v_{\text{opt}}\in\Omega_v$, where $v\in\mc V_i$, such that
    \begin{equation}
        \sum_{v\in\mc V_i}x^v_{\text{opt}}-n^i=0\,.
    \end{equation}
    Here, choosing vector $x^v_{\text{opt}}$ corresponds to choosing the charging station $j_{\text{opt}}^v$. For every vehicle $v$ and $j\neq j_{\text{opt}}^v$, we let $x^v_j\in\Omega_v\setminus\left\{x^v_{\text{opt}}\right\}$. For a particular choice $x^v_j$ of a charging station, the cost function of the driver can be simplified to
    \begin{equation}
        J^v\left(x^v_j,\rho^v\right)=\alpha^v_j-\left(\beta^v_j\right)^T\rho^v\,,
    \end{equation}
    for some $\alpha^v_j\in\R$ and $\beta^v_j\in\R^{\mc M}$. Notice from the structure of the loss function of each vehicle that for each $j$, the vector $\beta^v_j$ has all entries equal to 0 except for the $j$-th component which is positive. The vector $x^v_{\text{opt}}$ will be a solution of the problem~\eqref{mini:op1} if for every $j\in\mc M\setminus\left\{j_{\text{opt}}^v\right\}$ it holds that
    \begin{equation}
        \alpha^v_j-\left(\beta^v_j\right)^T\rho^v\geq\alpha^v_{\text{opt}}-\left(\beta^v_{\text{opt}}\right)^T\rho^v\,.
        \label{eq:m1}
    \end{equation}
    If $\overline{b}_j>0$ denotes the $j$-th element of vector $\beta^v_j$ and $\overline{\delta}_j=\alpha^v_j-\alpha^v_{\text{opt}}$, then rearranging~\eqref{eq:m1} gives an equivalent system of $m_s-1$ linear inequalities
    \begin{equation}
        \overline{b}_{j_{\text{opt}}^v}\left(\rho^v\right)_{j_{\text{opt}}^v}\geq\overline{b}_j\left(\rho^v\right)_j-\overline{\delta}_j\,.
        \label{eq:m2}
    \end{equation}
    This system always has a solution such that $\rho^{v}\geq \rho_{\text{min}}$ as we can arbitrarily choose all $\left(\rho^{v}\right)_{j}\geq\left(\rho_{\text{min}}\right)_{j}$ and pick the $\left(\rho^{v}\right)_{j_{\text{opt}}^v}$ component such that the following holds:
    \begin{equation}
        \left(\rho^{v}\right)_{j_{\text{opt}}^v}\geq\frac{1}{\overline{b}_{j_{\text{opt}}^v}}\max\left\{\left(\rho_{\text{min}}\right)_{j_{\text{opt}}^v}, \displaystyle\max_{j}\left\{\overline{b}_j\left(\rho^v\right)_j-\overline{\delta}_j\right\}\right\}\,.
    \end{equation}
    This means that it is always possible to construct a vector $\rho^v$ such that $x^v_{\text{opt}}$ is part of the solution of~\eqref{mini:op1}, which in return yields an optimal value  $J^i_M\left(\sigma\left(\mu^{i*}\right) |\: n^i\right)=0$.  
\end{proof}
\medskip

Note that in this setup the company operator does not place any constraints on the surge price distribution among the vehicles. The operator could be interested in maintaining absolute fairness in a sense that all the vehicles should receive the same surge prices. Therefore, the operator could impose a constraint that the surge prices be equal, i.e., for every $v_1,v_2\in\mc V_i$ such that $v_1\neq v_2$, it should hold that $\rho^{v_1}=\rho^{v_2}$. This would correspond to a centralized optimization problem~\eqref{mini:op2}
\begin{mini}
{\rho,x^{v}, v\in\mc V_{i}}{J^i_M\left(\sigma\left(\mu^i\right) |\: n^i\right)}
{}{}
\addConstraint{x^{v}\in\argmin_{\tilde{x}^{v}\in\Omega_{v}}J^{v}\left(\tilde{x}^{v},\rho\right), \forall v\in\mc V_{i}}{}
\addConstraint{\rho\geq \rho_{\text{min}}}{}
\addConstraint{x^{v}\in\Omega_{v}, \forall v\in\mc V_{i}}{}
\addConstraint{\mu^i=\sum_{v\in\mc V_i}x^v}{}
\label{mini:op2}
\end{mini}
However, under this constraint, the $N_i$ optimality conditions $x^{v}\in\argmin_{\tilde{x}^{v}\in\Omega_{v}}J^{v}\left(\tilde{x}^{v},\rho\right)$ will no longer result in a set of $N_i$ uncoupled systems of linear inequalities given by~\eqref{eq:m1}. In fact, they will result in a system of $N_im_s$ linear inequalities whose parameters directly depend on the state of the fleet and might in certain scenarios lead to having the optimal value of~\eqref{mini:op2} greater than 0.
\begin {figure*}
\centering
\begin{adjustbox}{max height=0.45\textwidth, max width=\textwidth}
     \begin{subfigure}
         \centering
         \input{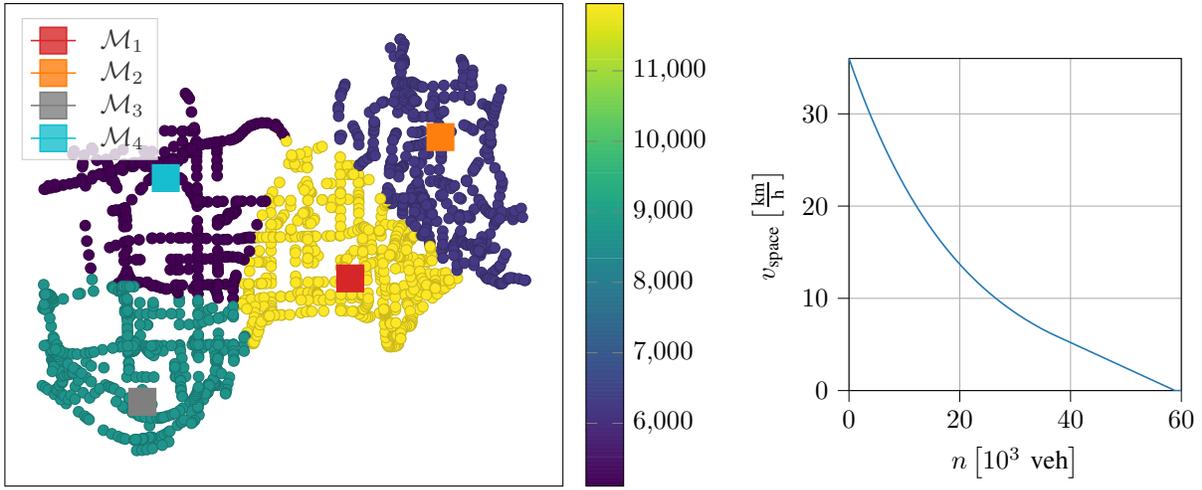}
         \label{fig:Shenzhen}
     \end{subfigure}
     \begin{subfigure}
         \centering
         \input{figures/MFD.tikz}
         \label{fig:MFD}
     \end{subfigure}
\end{adjustbox}
\caption{Map of Shenzhen - The network topology used in the case study consists of 1858 intersections connected by 2013 road segments divided in 4 regions around charging stations $\mc M=\left\{\mc M_1,\mc M_2,\mc M_3,\mc M_4\right\}$ according to the Voronoi partitioning of the city. Color of the nodes within each region indicates the total number of ride-hailing requests whose origin is in that region. From the left part of the figure, it is clear that the highest number of ride-hailing requests occurs in the region around charging station $\mc M_1$ and the lowest number of requests occurs in the region around $\mc M_4$. The right part of the figure shows the macroscopic fundamental diagram (MFD) for the city of Shenzhen according to \cite{doi:10.3141/2422-01}.}
\label{fig:case_study}
\end{figure*}
With this in mind, we propose that each company $i\in\mc C$ performs a two-step optimization procedure to determine the optimal surge prices. The first step consists of solving the optimization problem~\eqref{mini:op2} which ensures all drivers receive the same surge prices. If the optimal $\rho^*$ yields $J^i_M\left(\sigma\left(\left(\mu^i\right)^*\right) |\: n^i\right)>0$, then the second step that consists of solving~\eqref{mini:op1} is applied. This will ensure $J^i_M\left(\sigma\left(\left(\mu^i\right)^*\right) |\: n^i\right)=0$ at the expense of utilizing different surge prices for different vehicles. 

In the following section we will introduce in detail a case study based on taxi data from the city of Shenzhen. We will demonstrate the performance of both control levels and illustrate the robustness of the algorithm.

\section{Case study}\label{sec:example}

 We consider 3 ride-hailing companies $\mc C=\left\{\mc C_1, \mc C_2, \mc C_3\right\}$ with fleet sizes given by $N_{\text{fleet}}=\left[450, 400, 350\right]^{T}$ that operate in the Shenzhen region with 4 public charging stations $\mc M=\left\{\mc M_1, \mc M_2, \mc M_3, \mc M_4\right\}$. The stations are described by the vector of their capacities $M=\left[15, 60, 35, 50\right]^{T}$ and are located in parts of Shenzhen with different demands for ride-hailing services as shown in Figure \ref{fig:case_study}. We consider a 3 hour long simulation that represents one of the two peak-hour periods during the day. New passengers constantly arrive in the system and either increase the number of private vehicles in the system or request a ride-hailing vehicle to be assigned to them. The demand profile represents the real taxi demand that we assume is now served by the ride-hailing companies~\cite{BEOJONE2021102890}. The passengers are matched with the vehicles such that the total number of served requests is maximized, the total waiting time of passengers is minimized and no passenger has to wait for more than $\Delta\tau=10\:\text{min}$ for the ride-hailing vehicle to arrive. If the passenger is not matched with a taxi, he will opt for using the private vehicle. To better represent the congested conditions, the space mean speed of the vehicles is modelled as a decreasing function of the total vehicle accumulation $n$ in the region and according to the network Macroscopic
Fundamental Diagram (MFD) \cite{GEROLIMINIS2008759} obtained from \cite{doi:10.3141/2422-01}.  Under the assumption of homogeneous congestion in the city, the MFD of the region is given by:
\begin{equation}
   v_{\text{space}}(n)= \begin{cases}36 \exp{\left(-\frac{29n}{60000}\right)}, & \text { if } \frac{n}{1000} \leq 36 \\ 6.31-0.28\left(\frac{n}{1000}-36\right), & \text { if } 36<\frac{n}{1000} \leq 60 \\ 0, & \text { if } \frac{n}{1000}>60\end{cases}\,. 
\end{equation}
To prevent the ride-hailing vehicles from flocking in the busiest parts of the city, the desired distribution of the ride-hailing vehicles $\mc Z$ is formed so as to match the spatial distribution of the ride-hailing service requests. To approximate this distribution, the city region is divided into 4 cells according to the Voronoi partitioning of the map~\cite{Kang2008}. The charging stations are chosen as the centroids of the Voronoi cells and $\mc Z$ is chosen so as to correspond to the total number of requests in each cell. 

For every company $i\in \mc C$, the state of each vehicle $v \in \mc V_{i}$ to be charged is described by a tuple $\left(s_{v}^{\text{curr}}, s_{v}^{\text{des}}, d_{v}^{\text{max}}\right)$ where $s_{v}^{\text{curr}}$ represents the current battery level in percentage, $ s_{v}^{\text{des}}$ is set to $100\%$ and $d^{\text{max}}_v$ is the maximal range of the vehicle. We assume that each vehicle starts the simulation with a battery level chosen uniformly at random between $90\%$ and $95\%$. After 3 hours of operation, the ride-hailing vehicles whose battery level dropped below their personal threshold $t_v$ opt for charging. For each vehicle $v$, the threshold is sampled from a uniform distribution $t_v\sim \mathcal{U}[55, 60]$. For simplicity, we assume a linear discharge model of the battery given by
\begin{equation}
    s^{\text{curr}}_v\left(t+\Delta T\right)=s^{\text{curr}}_v\left(t\right)-\frac{100}{d^{\text{max}}_v}v_{\text{space}}\left(t\right)\Delta T\,.
\end{equation}
 A charging station $k$ is considered to be feasible for vehicle $v$ if it is within reach given the current battery status, i.e., if $s_{v}^{\text{curr}}-\frac{100}{d_{v}^{\text{max}}}d_{v,k}>0$ where $d_{v,k}$ denotes the shortest path between the vehicle $v$ and the charging station $k$. For every company $i\in\mc C$, the charging cost~\eqref{eq:J2} is modelled by setting $D_{i}\defineas N_{i}R_{i}$ where diagonal matrix $R_{i}\in \mathbb{R}^{4\times 4}$ captures the average charging demand per vehicle when choosing each of the charging stations. If a station is infeasible, the average demand is set to 0. Conversely, if the charging station $k$ is feasible to vehicle $v\in \mc V_{i}$, then vehicle's charging demand if $k$ is chosen for charging is defined as 
 \begin{equation}
     \delta_{v,k}=\beta_{v}\left(s_{v}^{\text{des}}-\left(s_{v}^{\text{start}}-\frac{100}{d_{v}^{\text{max}}}d_{v,k}\right)\right)\,,
 \end{equation}
where $\beta_{v}\in \mathbb{R}$ is a scaling coefficient that tells how many units of charge corresponds to $1\%$ of the vehicle's battery. The diagonal element of $R_{i}$ that corresponds to station $k$ is then given by
\begin{equation}
   (R_{i})_{kk}=\frac{1}{\left|\mc F_{k}^{i}\right|}\sum_{v\in \mc F_{k}^{i}}\delta_{v,k}\,.
\end{equation}  

The term that describes the negative expected revenue in~\eqref{eq:companyloss} is given by setting
\begin{equation}
    f_{i}\defineas N_{i}\left(e^{\text{arr}}_{i}-e^{\text{pro}}_{i}\right)\,,
\end{equation}
where the parameter $e_{i}^{\text{arr}}\in \mathbb{R}^{\mc M}$ represents the average cost of a vehicle being unoccupied while traveling to a charging station, and the parameter $e_{i}^{\text{pro}}\in \mathbb{R}^{\mc M}$ denotes the expected profit in regions around different charging stations estimated from historical data. If station $k$ is infeasible for company $i\in\mc C$, then we set $\left(e_{i}^{\text{arr}}\right)_{k}=0$, otherwise it is equal to: 
\begin{equation}
    \left(e_{i}^{\text{arr}}\right)_{k}=u_{i}\cdot P_{k}\cdot\left[\frac{1}{\left|\mc F_{k}^{i}\right|}\sum_{v\in \mc F_{k}^{i}}d_{v,k}\right]\,,
\end{equation}
where $u_{i}\in \mathbb{R}$ is the monetary value of a vehicle being occupied while driving for $1 \text{ km}$, given in $\left[\$/\text{km}\right]$ and $P_{k}$ is the probability of a vehicle being occupied in the region around charging station $k$. We also set the vector $e_{i}^{\text{pro}}\in \mathbb{R}^{3}$ according to the desired distribution $\mc Z$ such that
\begin{equation}
    \left(e_{i}^{\text{pro}}\right)_{k}=\varphi\cdot\left(\mc Z\right)_{k}+w_k\,,
\end{equation}
where $\varphi\in\R_+$ is a scaling coefficient and $w_k\sim \mathcal{U}[-10, 10]$. 

For every vehicle $v\in\mc V_i$, the parameters $D_v$, $g_v$ and $H_v$ describe driver's loss function~\eqref{eq:costdriver} when choosing which station to pick. The diagonal element of $D_v$ that corresponds to station $k$ is given by $\left(D_v\right)_{kk}=\delta_{v,k}$. If we assume the expected profit $e_{i}^{\text{pro}}$ is estimated for a work shift of $T_{\text{daily}}$ hours, then the parameter vector $g_v$ that describes the standard negative expected revenue for operating $\tau_v$ hours in the regions around charging stations is given by
\begin{equation}
    g_v=e^{\text{arr}}_{i}-\frac{\tau_v}{T_{\text{daily}}}e^{\text{pro}}_{i}\,.
\end{equation}
The diagonal entries of the matrix $H_v$ that describe the additional profit due to surge pricing are given by
\begin{equation}
    \left(H_v\right)_{kk}=\tau_v\cdot\overline{v}_{\text{space}}\cdot P_k\,,
\end{equation}
where $\overline{v}_{\text{space}}\in\R_+$ is the driver's estimate of the space mean speed in the network.

In this particular scenario, we fix the values of other parameters to $\beta_{v}=1.0$, $Q=0.1\cdot\text{diag}(4, 1, 3,2)$, $\varphi=300.0$ and $A_{G}=2.5Q$, vector of probabilities of being occupied $P=[0.35, 0.1, 0.2, 0.15]$, $u_{i}=1.0, \forall i\in\mc C$, $\tau_v=2$ for all vehicles, $T_{\text{daily}}=8$, $\overline{v}_{\text{space}}=20.0$ and set the number of iterations for the upper level control algorithm to $k=1000$.

\subsection{Numerical results -- upper-level control}
After a 3 hour simulation, the total number of vehicles that need to go and recharge is given by $N=\left[194,181,157\right]^T$. In order to illustrate the merits of using the reverse Stackelberg pricing mechanism, we will compare it against two baselines: 
\begin{itemize}
    \item A scenario in which the government does nothing, i.e., sets the same prices of charging for all stations. In this particular scenario, we set the baseline prices to $\overline{p}_{\text{base}}=\left[3.0,3.0,3.0,3.0\right]^T$.
    \item A scenario in which the prices of charging at each station can be different but are the same for all companies, i.e., for every station $k$ and any two $i,j\in\mc C$, it holds that $\left(p_i\left(x^i,x^{-i}\right)\right)_k=\left(p_j\left(x^j,x^{-j}\right)\right)_k$.
\end{itemize}
The two baseline scenarios essentially correspond to playing a Stackelberg game between the government and the ride-hailing companies. Analogous to the proof outlined in Theorem~\ref{th:t2}, since for all $i\in\mc C$ the matrix $A_i$ is positive semi-definite, the Stackelberg game played for any fixed pricing vector $\overline{p}\in\R_+^{\mc M}$ will have a unique Nash equilibrium. The second scenario is a more intuitive pricing approach as the price of charging depends only on the choice of charging station. However, it experiences reduced flexibility in terms of minimizing the central operator's objective and additional computational complexity. Namely, in order to compute the optimal pricing vector $\bar{p}\in\overline{\mc P}\subseteq\R^{\mc M}$ for the Stackelberg-based mechanism, we would in general have to solve 
\begin{equation}
    \bar{p}^*= \argmin_{p\in\overline{\mc P}}\frac{1}{2}\sigma\left(x^*\left(\bar{p}\right)\right)^{T}A_{G} \sigma\left(x^*\left(\bar{p}\right)\right)+b_{G}^{T}\sigma\left(x^*\left(\bar{p}\right)\right)\,,
    \label{eq:stackelbergp}
\end{equation}
where $x^*\left(\bar{p}\right)\in \mc X$ is the Nash equilibrium of the game \eqref{eq:gameform} for a particular fixed price choice $\bar{p}$. However, the closed form $x^*\left(\bar{p}\right)$ is in general not known. Instead, in order to approximate the optimal $\bar{p}$, we choose a fixed maximum price $p_{\text{max}}\in\R_+$, set $\overline{\mc P}=\left[0,p_{\text{max}}\:\right]^{\mc M}$ and perform extensive local grid search. 
\begin{figure}[tbp]
    \centering
    \input{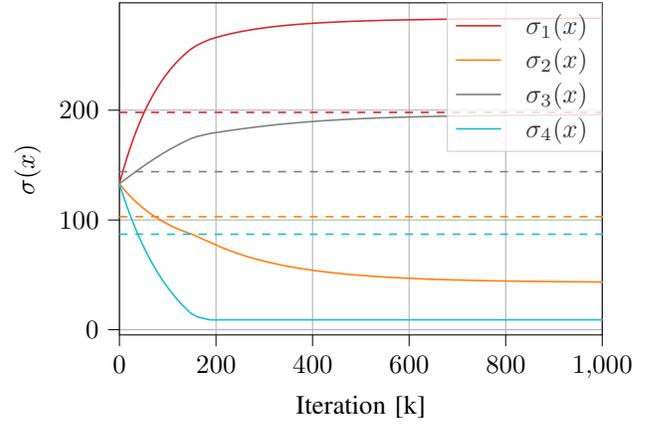}
    \caption{Evolution of the total number of vehicles at each charging station when no action is taken, i.e., $p=p_{\text{base}}=\left[3.0,3.0,3.0,3.0\right]^T$. The dashed line represents the desired value of the vehicle accumulation whereas the solid line represents the attained value.}
    \label{fig:noaction}
\end{figure}

Figure~\ref{fig:noaction} shows the attained Nash equilibrium of the game when no action is taken, i.e., for every $i\in\mc C$ we use $p_i\left(x^{i},x^{-i}\right)=p_{\text{base}}=\left[3.0,3.0,3.0,3.0\right]^T$. As expected, the regions around the charging stations $\mc M_1$ and $\mc M_3$, that are more popular in terms of demand, will attract more ride-hailing vehicles when the prices of charging are the same for different stations. This, in return, results in very small occupancy of the charging station $\mc M_4$ that is located in the region with the smallest number of ride-hailing requests and hence provides the smallest expected profit.
\begin{figure}[tbp]
    \centering
    \input{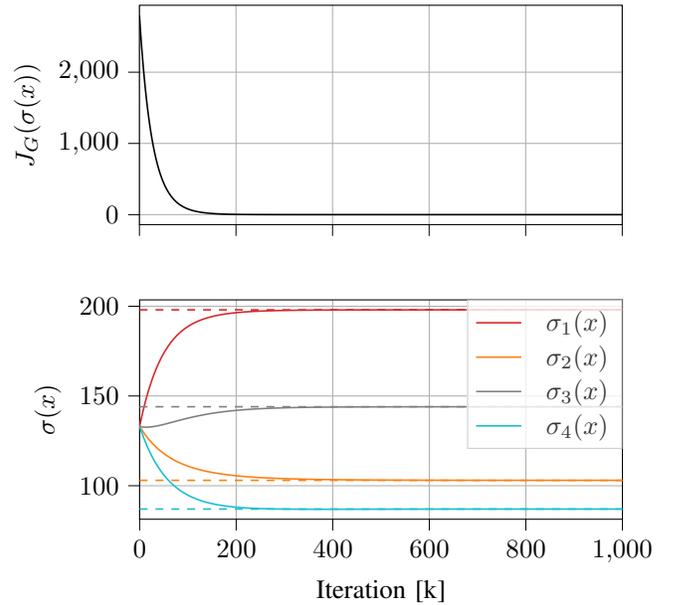}
    \caption{Performance of the reverse Stackelberg game pricing mechanism. The upper subplot shows the convergence of the government's loss function whereas the lower subplot shows the evolution of the total number of vehicles at each charging station. Here, $\sigma_j\left(x\right)$ corresponds to the total number of vehicles at charging station $\mc M_j$, dash line corresponds to desired and solid line to attained value of the total number of vehicles.}
    \label{fig:rsg1}
\end{figure}

When system optimal pricing policies in Definition~\ref{def:pricingpolicies} are used, the distribution of the car fleets over the charging stations and the resulting charging
prices are presented in Table~\ref{tab:1} whereas the evolution of the government loss $J_G$ and the total number of vehicles over the iterations is presented in Figure~\ref{fig:rsg1}.
\begin{table}
\begin{center}
 \renewcommand{\arraystretch}{1.2}
 \caption{Company decisions and charging prices}\vspace{1ex}
 \label{tab:1}
 \begin{tabular}{c|cc|cc|cc|cc}
 \multirow{2}{*}{} & \multicolumn{2}{c|}{Station $\mc M_1$} & \multicolumn{2}{c|}{Station $\mc M_2$} & \multicolumn{2}{c|}{Station $\mc M_3$} & \multicolumn{2}{c}{Station $\mc M_4$} \\ 
 & $x_{1}^{i}$ & $p_{i}$ &$x_{2}^{i}$ & $p_{i}$ & $x_{3}^{i}$ & $p_{i}$ & $x_{4}^{i}$ & $p_{i}$ \\
 \hline

 $\mc C_{1} \rule{0pt}{2.6ex}$ &  0.38 & 3.99  & 0.19 & 3.00  & 0.27 & 3.54  & 0.16 & 2.37  \\
 $\mc C_{2}$ &  0.37 & 4.06  & 0.20 & 2.99  & 0.27 & 3.57  & 0.16 & 2.38 \\
 $\mc C_{3}$ &  0.36 & 4.22  & 0.20 & 3.03  & 0.27 & 3.64  & 0.17 & 2.41 \\
 \hline
   $\hat{N}$& \multicolumn{2}{c|}{198} & \multicolumn{2}{c|}{103} & \multicolumn{2}{c|}{144} & \multicolumn{2}{c}{87}\\
\hline
  $\sigma\left(x^*\right)$& \multicolumn{2}{c|}{198} & \multicolumn{2}{c|}{103} & \multicolumn{2}{c|}{144} & \multicolumn{2}{c}{87}\\
\hline 
\end{tabular}
\end{center}
\end{table}
From the figure it is clear that the iterative procedure converged to a Nash equilibrium that is simultaneously the global optimum of the government's objective and that attains $J_G\left(\sigma\left(x^*\right)\right)=0$. It is noteworthy to mention here that the vehicle configuration tested in this scenario facilitated the possibility to obtain the minimal possible value of $J_G$. As shown in~\cite{9838005}, it is just as probable that the battery level and the position of the ride-hailing vehicles, encoded in constraint sets $\mc X_i$, will not allow to perfectly match desired vehicle accumulations around different charging stations. As expected, the prices of charging are higher for the two most demand attractive regions, i.e., the ones centered at $\mc M_1$ and $\mc M_3$. Station $\mc M_4$ is the least attractive hence, it has the smallest charging prices in the Nash equilibrium. Looking at the prices at a particular station, it is clear that the good performance of the system when minimizing the government's objective comes at the expense of utilizing different prices of charging at a particular station for different ride-hailing companies. Because there is no constraint on how big the price difference for individual companies can be, there is a risk of having unfair prices for certain companies. Though this is not the case in the displayed scenario, it is important to theoretically examine in the future the question of fair prices across the companies.
\begin{table}
\begin{center}
 \renewcommand{\arraystretch}{1.2}
 \caption{Comparison of different pricing mechanism}\vspace{1ex}
 \label{tab:2}
 \begin{tabular}{c|c|cccc}
 \multirow{2}{*}{}  & \multirow{2}{*}{$J_G$} & $\sigma_1\left(x^*\right)$ & $\sigma_2\left(x^*\right)$ & $\sigma_3\left(x^*\right)$ & $\sigma_4\left(x^*\right)$ \\
 & & $\widehat{N}_1=198$ & $\widehat{N}_2=103$ & $\widehat{N}_3=144$ & $\widehat{N}_4=87$ \\
 \hline
 $\overline{p}_{\text{base}}$         &  6677.9 & 283.9  &  43.03 & 196.0 & 8.999  \\
 $\overline{p}_{1}$                   &  13.585 & 200.8  &  98.43 & 147.9 & 84.81   \\
 $\overline{p}_{2}$                   &  18.579 & 198.2  &  111.1 & 140.2 & 82.49   \\
 RSG                                &  0.0000 & 198.0  &  103.0 & 144.0 & 87.00   \\
 \hline
\end{tabular}
\end{center}
\end{table}

The complete performance comparison of the system using the system optimal pricing policies given in Definition~\ref{def:pricingpolicies} against the Stackelberg based pricing mechanisms with $\overline{p}_{\text{base}}$, $\overline{p}_1=\left[2.75, 1.625, 2.208, 1.0\right]$ and $\overline{p}_2=\left[4.03, 2.8, 3.49, 2.24\right]$ is given in Table~\ref{tab:2}. Values $\overline{p}_1$ and $\overline{p}_2$ are obtained via local grid search. They attain government's losses $J_G$ in the Nash equilibrium that are close in value making them both a viable choice for charging prices. It is evident that the reverse Stackelberg-based pricing mechanism outperforms every other baseline scenario in terms of minimizing the government's loss function. However, in light of the robustness analysis presented in Section~\ref{subsec:robust}, we next compare the performance of the system optimal pricing policies against the two Stackelberg-based pricing mechanisms $\overline{p}_1$ and $\overline{p}_2$ when different levels of uncertainty are introduced in the parameter $D_i$.
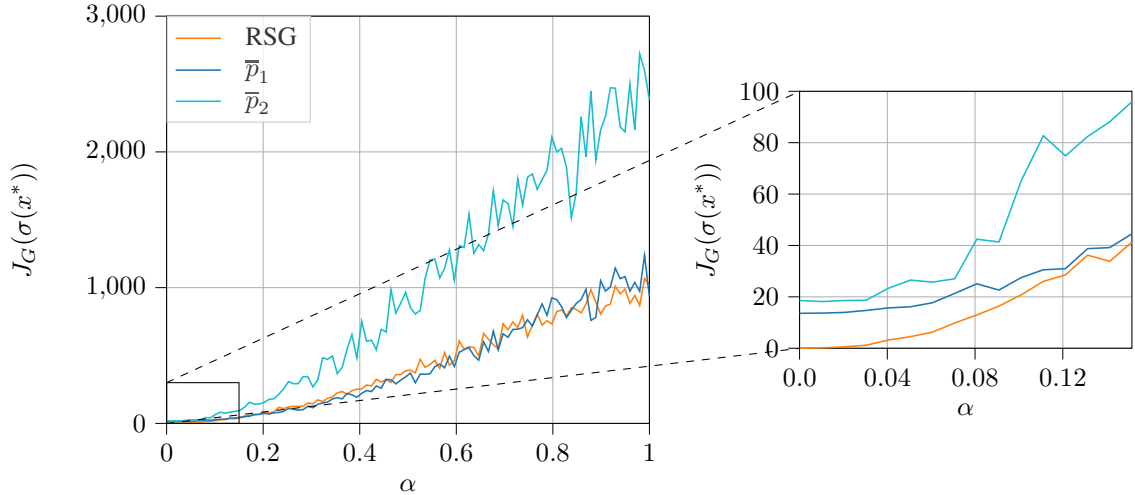
\begin {figure*}
\centering
\begin{adjustbox}{max height=0.45\textwidth, max width=\textwidth}
    \input{figures/robust.tikz}
\end{adjustbox}
\caption{Robustness plot for different levels of perturbation magnitude $\alpha$. For every $\alpha$, we sample $w_{k}$ 100 times and report the mean value of the government's loss in the Nash equilibrium when the system optimal pricing policies, $\overline{p}_1=\left[2.75, 1.625, 2.208, 1.0\right]$ and $\overline{p}_2=\left[4.03, 2.8, 3.49, 2.24\right]$ are applied.}
\label{fig:rob1}
\end{figure*}

According to the system structure used in this case study, apart from $R_{i}$ and $e^{\text{arr}}_i$, all other parameters are inherently known to the government as they characterize the available charging infrastructure and the dynamics of the ride-hailing market in the region where the companies operate. As previously noted, the government optimum is attainable if the companies are willing to share $R_{i}$ and $e^{\text{arr}}_i$ that encompass the information about the average charging demand and position of the company's fleet. $R_{i}$ is considered a more sensitive piece of information compared to $e^{\text{arr}}$ as the latter one only depends on the vehicles' distances to charging stations. Hence, we show how the system optimal pricing mechanism behaves in the same scenario when the government has only an estimate $\overline{D}_{i}$ of the average charging demand $D_{i}= N_iR_i$. For a feasible station $k$, i.e., $\left(D_{i}\right)_{kk}>0$, we let $\left(\overline{D}_{i}\right)_{kk}=\left(D_{i}\right)_{kk}+w_{k}$ where $w_{k}$ is a noise sample drawn from a normal distribution such that $w_{k}\sim \mathcal{N}\left(0,\left(\alpha D_{\text{min}}/4\right)^2\right)$ with $D_{\text{min}}$ being the minimal, non-zero, diagonal element of $D_{i}$. For every magnitude of uncertainty $\alpha$, we sample $w_{k}$ 100 times and report the mean value of the government's loss in the Nash equilibrium. The corresponding plots are shown in Figure~\ref{fig:rob1}. For moderate discrepancies $\left(\alpha<0.15\right)$ between the true and the estimated value of $D_{i}$, the attained Nash equilibrium with system optimal pricing policies is close to the government's optimum. Moreover, the enlarged part of Figure~\ref{fig:rob1} suggests that utilizing system optimal pricing policies results in a more robust mechanism with respect to parameter $D_i$. 

It is particularly interesting to analyze what happens for larger discrepancies, i.e., $\alpha>0.15$. Based on the local minimum in which the minimization procedure~\eqref{eq:stackelbergp} lands, we observe significantly different robustness characteristics of the system when the magnitude of the perturbation is increased. Even though both Stackelberg prices $\overline{p}_1$ and $\overline{p}_2$ yield similar government objective values when parameter $D_i$ is perfectly known, using $\overline{p}_2$ seems to significantly underperform compared to the system optimal pricing policies for any magnitude of the perturbation $\alpha$. Conversely, the prices $\overline{p}_1$ on average match the performance of the system optimal pricing policies. It is important to note here that in order to compute the local optima of~\eqref{eq:stackelbergp} we only performe an extensive grid search procedure over the space of charging price vectors defined by $\overline{\mc P}$. Without a heuristic and a verification method to help us find the local optima for which the robustness with respect to $D_i$ can be verified, using a local search to approximate the optimal $\overline{p}$ can result in poor robustness performance.  

It is worth mentioning that the execution time of all simulations with system optimal pricing policies on an average PC was smaller than 3 sec, which substantiates the possibility of applying the algorithm to a real system.

\begin {figure*}
\centering
\begin{adjustbox}{max height=0.45\textwidth, max width=\textwidth}
    \input{figures/hist.tikz}
\end{adjustbox}
\caption{Distribution of surge prices used by the ride-hailing companies for different charging stations. With a slight abuse of notation, the label $\rho^v$ on the x-axis denotes a single real number, not the whole vector of surge prices for a particular vehicle $v\in\mc V_i$. For clarity, the surge price of zero has been removed from the plot as there is a significant number of vehicles that do not need to be proposed a surge price for a particular station in order to match the company operator's desired $n^{i}$.}
\label{fig:hist}
\end{figure*}
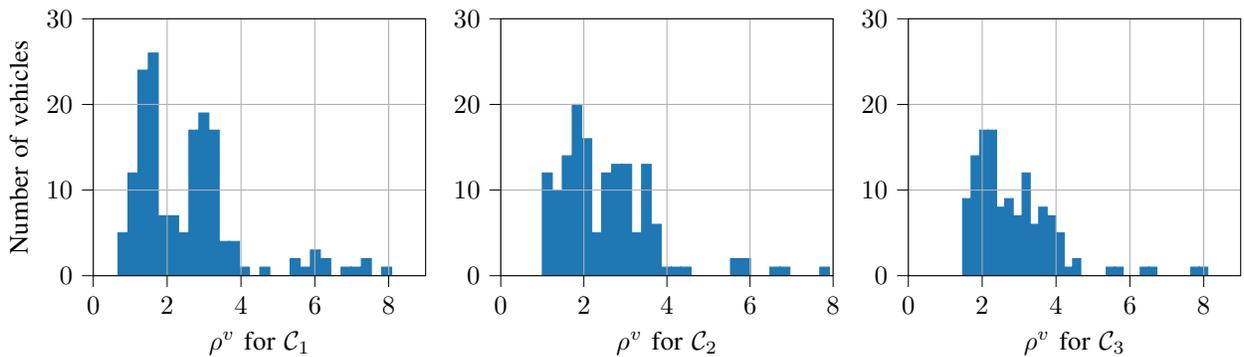

\subsection{Numerical results -- lower-level control}

Finally, we present the results of using the surge pricing mechanism to motivate the drivers $v\in\mc V_i$ of a particular company $i\in\mc C$ to follow the company operator's desired value of the discrete allocation vector $n^{i}$. For the ride-hailing fleet configurations used in this study, it was required to perform the complete two-step mixed-integer optimization procedure proposed in Section ~\ref{sec:matching}. With a slight abuse of notation, if $\rho^v\in\R_+$ denotes the value of a surge price proposed for a region around a single charging station and not the complete surge price vector $\rho^v\in\R^{\mc M}_+$ offered to a particular driver $v\in\mc V_i$, then the distribution of non-zero surge prices used by each company operator is given in Figure~\ref{fig:hist}. We observe that for all three ride-hailing companies, most of the proposed surge prices are localised in the lower end of the price spectrum. However, there are also vehicles that need to be offered large surge prices for certain regions which in return can result in the cancellation of the ride-hailing request. A naive way to overcome this would be to introduce an empirically determined upper bound on the surge price. However, similarly to making surge prices equal per vehicle, imposing an upper bound on the surge price could result in optimal value $J_M^{i}>0$.    

\section{Conclusion}\label{sec:conclusion}

In this work, we developed a bi-level model for balancing the demand of electric ride-hailing fleets on the charging stations in a particular region at a particular point in time, after a peak-hour service period during the day. On the upper control level, our approach is a reverse Stackelberg game between a central authority interested in optimizing a personal objective such as balancing the load on the power grid, reducing congestion, etc., and a set of companies in the ride-hailing market. It is a one-leader, multiple-followers game firstly introduced in~\cite{9838005} that we adapt for a case study based on the real taxi data from the city of Shenzhen. 

Apart from theoretically analyzing the robustness of the upper-level control algorithm in this paper, we also 
compare its performance against three baseline scenarios with fixed prices per company. One scenario corresponds to having the same prices of charging at each station whereas the other two baselines are obtained based on the local computation of the optimal leader strategy in Stackelberg games. We illustrate that with a naive local search, we can obtain a Stackelberg-based pricing mechanism that significantly underperforms in terms of robustness with respect to the parameter that describes the total charging demand of a company to be served at particular charging stations. Finally, on the lower control level, we introduce a Stackelberg-based surge pricing mechanism to provide financial incentives to drivers within a company in order to make them follow the company operator's desires. 

This work opens numerous research directions for the future. For instance, it is of the paramount importance to address the notion of fairness, be it among the ride-hailing companies on the upper level or among the drivers within a company on the lower level. Furthermore, increasing the complexity of the model by introducing coupling constraints, e.g., imposed by the power-grid, should be investigated in order to make the model more realistic. Finally, this work analyzes a static scenario in the sense that the state of the system is analyzed for only one particular point of time during the day. In the future, we plan to investigate if combining this framework with the ride-hailing demand prediction problem could result in a dynamic setup that would allow the companies to be proactive when planning when and how much to charge their fleets during the day.  



\bibliographystyle{IEEEtran}
\bibliography{references.bib}

\end{document}

%% file: figures/MFD.tikz
\begin{tikzpicture}

\definecolor{color0}{rgb}{0.12156862745098,0.466666666666667,0.705882352941177}

\begin{axis}[
tick align=outside,
tick pos=left,
x grid style={white!69.0196078431373!black},
xlabel={$n\left[10^3\text{ veh}\right]$},
xmajorgrids,
xmin=0, xmax=60,
xtick style={color=black},
y grid style={white!69.0196078431373!black},
ylabel={$v_{\text{space}}\left[\frac{\text{km}}{\text{h}}\right]$},
ymajorgrids,
ymin=0.0, ymax=36,
ytick style={color=black},
width = 6cm,
height = 6cm
]
\addplot [semithick, color0]
table {%
0 36
0.894089460372925 34.4774208068848
1.79417943954468 33.0096588134766
2.70627069473267 31.5860595703125
3.61836194992065 30.2238521575928
4.5424542427063 28.9036235809326
5.47254705429077 27.6330490112305
6.40864086151123 26.4106674194336
7.35073518753052 25.2350368499756
8.29883003234863 24.1047477722168
9.25892543792725 23.0117321014404
10.2310228347778 21.9555397033691
11.2091207504272 20.9417495727539
12.1992197036743 19.9631881713867
13.201319694519 19.0193176269531
14.2214221954346 18.1043128967285
15.2535257339478 17.2233352661133
16.3036308288574 16.3709774017334
17.3717365264893 15.5472688674927
18.4518451690674 14.7564420700073
19.549955368042 13.9936609268188
20.6720676422119 13.2549228668213
21.8121814727783 12.5442638397217
22.97629737854 11.8579425811768
24.1644172668457 11.196174621582
25.376537322998 10.5590810775757
26.6186618804932 9.94381046295166
27.8907890319824 9.35082054138184
29.1989192962646 8.77790546417236
30.5430545806885 8.22576522827148
31.9231929779053 7.69495296478271
33.3453330993652 7.18379497528076
34.8094825744629 6.69298982620239
36.2916297912598 6.23101663589478
58.8778877258301 0
60 0
};
\end{axis}

\end{tikzpicture}

%% file: figures/robust.tikz
\begin{tikzpicture}

\definecolor{color0}{rgb}{0.12156862745098,0.466666666666667,0.705882352941177}
\definecolor{color1}{rgb}{1,0.498039215686275,0.0549019607843137}
\definecolor{color2}{rgb}{0.0901960784313725,0.745098039215686,0.811764705882353}

\begin{axis}[
name=ax1,
legend cell align={left},
legend style={
  fill opacity=0.8,
  draw opacity=1,
  text opacity=1,
  at={(0.00,1.0)},
  anchor=north west,
  draw=white!80!black
},
tick align=outside,
tick pos=left,
width=8cm,
height=7cm,
x grid style={white!69.0196078431373!black},
xlabel={\(\displaystyle \alpha\)},
xmajorgrids,
xmin=0, xmax=1,
xtick style={color=black},
y grid style={white!69.0196078431373!black},
ylabel={\(\displaystyle J_G(\sigma(x^*))\)},
ylabel style={yshift=0.4cm},
ymajorgrids,
ymin=0.0, ymax=3000,
ytick style={color=black}
]
\addplot [semithick, color1]
table {%
0 0
0.0101009607315063 0.134645581245422
0.0202020406723022 0.554893970489502
0.0303030014038086 1.15439021587372
0.0404040813446045 3.17773747444153
0.0505050420761108 4.49404525756836
0.0606060028076172 6.26793050765991
0.0707070827484131 9.81321239471436
0.0808080434799194 12.9670219421387
0.0909091234207153 16.3944835662842
0.101010084152222 20.7692718505859
0.111111164093018 26.0110683441162
0.121212124824524 28.5582008361816
0.13131308555603 36.2324295043945
0.141414165496826 33.831787109375
0.151515126228333 41.2884750366211
0.161616206169128 46.4645957946777
0.171717166900635 60.0011749267578
0.181818246841431 62.0468139648438
0.191919207572937 66.7796020507812
0.202020168304443 78.9380340576172
0.212121248245239 76.7873687744141
0.222222208976746 84.7290573120117
0.232323169708252 118.045455932617
0.242424249649048 108.663902282715
0.252525210380554 122.500305175781
0.26262629032135 125.070541381836
0.272727251052856 126.260665893555
0.282828330993652 126.700523376465
0.292929291725159 149.906311035156
0.303030252456665 144.938705444336
0.313131332397461 173.0576171875
0.323232293128967 160.749313354492
0.333333373069763 184.880416870117
0.34343433380127 194.580581665039
0.353535413742065 225.064743041992
0.363636374473572 234.453659057617
0.373737335205078 214.672271728516
0.383838415145874 241.382476806641
0.39393937587738 251.692520141602
0.404040336608887 254.570159912109
0.414141416549683 282.672790527344
0.424242496490479 275.772186279297
0.434343457221985 325.163330078125
0.444444417953491 304.831024169922
0.454545497894287 322.540618896484
0.464646458625793 301.504180908203
0.4747474193573 359.578491210938
0.484848499298096 359.533264160156
0.494949460029602 387.333770751953
0.505050420761108 392.095031738281
0.515151500701904 372.144836425781
0.5252525806427 471.605346679688
0.535353541374207 413.015899658203
0.545454502105713 433.868286132812
0.555555582046509 464.84765625
0.565656542778015 405.44287109375
0.575757503509521 521.569152832031
0.585858583450317 428.60400390625
0.595959663391113 462.315948486328
0.60606062412262 606.544006347656
0.616161584854126 546.972351074219
0.626262664794922 564.259765625
0.636363625526428 559.351989746094
0.646464586257935 497.724578857422
0.65656566619873 662.68701171875
0.666666746139526 619.999450683594
0.676767587661743 578.722717285156
0.686868667602539 550.157531738281
0.696969747543335 734.412780761719
0.707070708274841 690.421752929688
0.717171669006348 746.817199707031
0.727272748947144 706.256225585938
0.73737370967865 795.507629394531
0.747474670410156 637.431701660156
0.767676830291748 804.642761230469
0.777777791023254 724.144836425781
0.787878751754761 755.835205078125
0.797979831695557 731.91162109375
0.808080792427063 823.266723632812
0.818181753158569 835.281127929688
0.828282833099365 791.311645507812
0.838383913040161 851.778686523438
0.848484873771667 839.326904296875
0.858585834503174 813.864868164062
0.86868691444397 880.445739746094
0.878787875175476 984.729431152344
0.898989915847778 806.075256347656
0.909090995788574 865.804565429688
0.919191837310791 878.445556640625
0.929292917251587 1009.84948730469
0.939393997192383 954.023193359375
0.949494957923889 990.114990234375
0.959595918655396 884.395751953125
0.969696998596191 942.144714355469
0.979797959327698 868.328369140625
0.989898920059204 1068.12475585938
1 1012.97241210938
};
\addlegendentry{RSG}
\addplot [semithick, color0]
table {%
0 13.5850591659546
0.0101009607315063 13.6772117614746
0.0202020406723022 13.9287672042847
0.0303030014038086 14.7013254165649
0.0404040813446045 15.6930360794067
0.0505050420761108 16.1205291748047
0.0606060028076172 17.7230110168457
0.0707070827484131 21.3227367401123
0.0808080434799194 25.0760631561279
0.0909091234207153 22.6413669586182
0.101010084152222 27.4850826263428
0.111111164093018 30.5344676971436
0.121212124824524 30.9653911590576
0.13131308555603 38.7574234008789
0.141414165496826 39.1793556213379
0.151515126228333 44.5317993164062
0.161616206169128 52.691722869873
0.171717166900635 52.3280143737793
0.181818246841431 66.0561294555664
0.191919207572937 66.8112258911133
0.202020168304443 74.2382431030273
0.212121248245239 68.8472366333008
0.222222208976746 77.7561416625977
0.232323169708252 80.8087768554688
0.242424249649048 88.3014373779297
0.252525210380554 110.756355285645
0.26262629032135 103.612525939941
0.272727251052856 97.7339172363281
0.282828330993652 101.864356994629
0.292929291725159 126.621940612793
0.303030252456665 115.483322143555
0.313131332397461 141.623901367188
0.323232293128967 150.539138793945
0.333333373069763 157.130630493164
0.34343433380127 179.810424804688
0.353535413742065 189.671813964844
0.363636374473572 180.714950561523
0.373737335205078 220.038284301758
0.383838415145874 210.759353637695
0.39393937587738 192.771087646484
0.404040336608887 221.136581420898
0.414141416549683 241.89225769043
0.424242496490479 234.396728515625
0.434343457221985 275.670288085938
0.444444417953491 259.084747314453
0.454545497894287 324.390930175781
0.464646458625793 321.308258056641
0.4747474193573 291.293640136719
0.484848499298096 318.374298095703
0.494949460029602 342.341949462891
0.505050420761108 370.800537109375
0.515151500701904 299.314483642578
0.5252525806427 363.126922607422
0.535353541374207 361.634918212891
0.545454502105713 364.940734863281
0.555555582046509 418.382049560547
0.565656542778015 437.078338623047
0.575757503509521 409.780059814453
0.585858583450317 494.706726074219
0.595959663391113 419.297790527344
0.60606062412262 524.247253417969
0.616161584854126 542.94384765625
0.626262664794922 558.689270019531
0.636363625526428 498.924926757812
0.646464586257935 465.643157958984
0.65656566619873 600.020141601562
0.666666746139526 529.884643554688
0.676767587661743 678.790832519531
0.686868667602539 542.251953125
0.696969747543335 660.18603515625
0.707070708274841 634.476379394531
0.717171669006348 692.413269042969
0.727272748947144 693.816772460938
0.73737370967865 705.937194824219
0.747474670410156 764.235168457031
0.757575750350952 716.517944335938
0.767676830291748 800.33447265625
0.777777791023254 880.180358886719
0.787878751754761 796.493530273438
0.797979831695557 924.848022460938
0.808080792427063 911.682861328125
0.818181753158569 859.431335449219
0.828282833099365 779.073059082031
0.838383913040161 834.370422363281
0.848484873771667 883.718627929688
0.858585834503174 912.371398925781
0.86868691444397 987.549743652344
0.878787875175476 759.010070800781
0.888888835906982 780.651062011719
0.898989915847778 1050.98779296875
0.909090995788574 974.587036132812
0.919191837310791 1005.82873535156
0.929292917251587 1143.38134765625
0.939393997192383 1066.88256835938
0.949494957923889 1081.27758789062
0.959595918655396 973.294738769531
0.969696998596191 1038.98059082031
0.979797959327698 973.11669921875
0.989898920059204 1237.65698242188
1 934.646667480469
};
\addlegendentry{$\overline{p}_1$}
\addplot [semithick, color2]
table {%
0 18.579963684082
0.0101009607315063 18.1745681762695
0.0202020406723022 18.5703773498535
0.0303030014038086 18.6628265380859
0.0404040813446045 23.3074188232422
0.0505050420761108 26.517505645752
0.0606060028076172 25.7311401367188
0.0707070827484131 27.0324268341064
0.0808080434799194 42.4336776733398
0.0909091234207153 41.3424797058105
0.101010084152222 65.0961990356445
0.111111164093018 82.7074584960938
0.121212124824524 74.8834838867188
0.13131308555603 82.3810043334961
0.141414165496826 88.137809753418
0.151515126228333 95.9365615844727
0.161616206169128 119.485328674316
0.171717166900635 156.448425292969
0.181818246841431 141.767868041992
0.191919207572937 143.512100219727
0.202020168304443 154.471282958984
0.212121248245239 176.887878417969
0.222222208976746 181.764541625977
0.232323169708252 215.136215209961
0.242424249649048 254.8740234375
0.252525210380554 308.289855957031
0.26262629032135 292.194580078125
0.272727251052856 293.084228515625
0.282828330993652 344.882080078125
0.292929291725159 341.822601318359
0.303030252456665 268.037292480469
0.313131332397461 319.833587646484
0.323232293128967 483.134307861328
0.333333373069763 471.547943115234
0.34343433380127 497.225616455078
0.353535413742065 366.454223632812
0.363636374473572 412.480834960938
0.373737335205078 546.57763671875
0.383838415145874 653.494567871094
0.39393937587738 483.470245361328
0.404040336608887 752.533386230469
0.414141416549683 604.686157226562
0.424242496490479 613.149719238281
0.434343457221985 543.340209960938
0.444444417953491 692.972778320312
0.454545497894287 672.700561523438
0.464646458625793 985.064758300781
0.4747474193573 811.761779785156
0.484848499298096 884.9462890625
0.494949460029602 965.451354980469
0.505050420761108 852.497741699219
0.515151500701904 805.783569335938
0.5252525806427 831.513671875
0.535353541374207 1067.45544433594
0.545454502105713 1204.14929199219
0.555555582046509 1210.96704101562
0.565656542778015 1072.2275390625
0.575757503509521 1142.07263183594
0.585858583450317 1277.68969726562
0.595959663391113 1054.7802734375
0.60606062412262 1309.72509765625
0.616161584854126 1296.17834472656
0.626262664794922 1540.31091308594
0.636363625526428 1252.51538085938
0.646464586257935 1317.57897949219
0.65656566619873 1272.11206054688
0.666666746139526 1396.34045410156
0.676767587661743 1713.35192871094
0.686868667602539 1458.00244140625
0.696969747543335 1646.03308105469
0.707070708274841 1619.20178222656
0.717171669006348 1451.17663574219
0.727272748947144 1800.79626464844
0.73737370967865 1608.404296875
0.747474670410156 1813.64123535156
0.757575750350952 1836.95812988281
0.767676830291748 1724.36279296875
0.777777791023254 1796.46911621094
0.787878751754761 1863.9013671875
0.797979831695557 2109.44213867188
0.808080792427063 1999.48876953125
0.818181753158569 2025.87622070312
0.828282833099365 1884.04296875
0.838383913040161 1516.51708984375
0.848484873771667 1701.55773925781
0.858585834503174 2259.39965820312
0.86868691444397 1964.84790039062
0.878787875175476 2448.94482421875
0.888888835906982 1938.84460449219
0.898989915847778 2163.30297851562
0.909090995788574 2264.015625
0.919191837310791 2472.00659179688
0.929292917251587 2469.13818359375
0.939393997192383 2185.33862304688
0.949494957923889 2148.51440429688
0.959595918655396 2499.0458984375
0.969696998596191 2163.78198242188
0.979797959327698 2722.03442382812
0.989898920059204 2599.6982421875
1 2385.72436523438
};
\addlegendentry{$\overline{p}_2$}

  \coordinate (c1) at (axis cs:0.0, 0.0);
  \coordinate (c2) at (axis cs:0.0, 300);
  \draw (c1) rectangle (axis cs:0.15,300);

\end{axis}

 \begin{axis}[
 xtick={0.0, 0.04, 0.08, 0.12},
xticklabels={%
$0.0$,
$0.04$,
$0.08$,
$0.12$,
$0.15$},
   name=ax2,
tick align=outside,
tick pos=left,
width=6cm,
height=5cm,
x grid style={white!69.0196078431373!black},
xlabel={\(\displaystyle \alpha\)},
xmajorgrids,
xmin=0, xmax=0.151515126228333,
ymin=0.0, ymax=100,
xtick style={color=black},
y grid style={white!69.0196078431373!black},
ylabel={\(\displaystyle J_G(\sigma(x^*))\)},
ymajorgrids,
ytick style={color=black}, 
   at={($(ax1.south east)+(2cm,1cm)$)},
 ]

\addplot [semithick, color1]
table {%
0 0
0.0101009607315063 0.134645581245422
0.0202020406723022 0.554893970489502
0.0303030014038086 1.15439021587372
0.0404040813446045 3.17773747444153
0.0505050420761108 4.49404525756836
0.0606060028076172 6.26793050765991
0.0707070827484131 9.81321239471436
0.0808080434799194 12.9670219421387
0.0909091234207153 16.3944835662842
0.101010084152222 20.7692718505859
0.111111164093018 26.0110683441162
0.121212124824524 28.5582008361816
0.13131308555603 36.2324295043945
0.141414165496826 33.831787109375
0.151515126228333 41.2884750366211
};%
\addplot [semithick, color0]
table {%
0 13.5850591659546
0.0101009607315063 13.6772117614746
0.0202020406723022 13.9287672042847
0.0303030014038086 14.7013254165649
0.0404040813446045 15.6930360794067
0.0505050420761108 16.1205291748047
0.0606060028076172 17.7230110168457
0.0707070827484131 21.3227367401123
0.0808080434799194 25.0760631561279
0.0909091234207153 22.6413669586182
0.101010084152222 27.4850826263428
0.111111164093018 30.5344676971436
0.121212124824524 30.9653911590576
0.13131308555603 38.7574234008789
0.141414165496826 39.1793556213379
0.151515126228333 44.5317993164062
};%
\addplot [semithick, color2]
table {%
0 18.579963684082
0.0101009607315063 18.1745681762695
0.0202020406723022 18.5703773498535
0.0303030014038086 18.6628265380859
0.0404040813446045 23.3074188232422
0.0505050420761108 26.517505645752
0.0606060028076172 25.7311401367188
0.0707070827484131 27.0324268341064
0.0808080434799194 42.4336776733398
0.0909091234207153 41.3424797058105
0.101010084152222 65.0961990356445
0.111111164093018 82.7074584960938
0.121212124824524 74.8834838867188
0.13131308555603 82.3810043334961
0.141414165496826 88.137809753418
0.151515126228333 95.9365615844727
};%

\end{axis}

\draw [dashed] (c1) -- (ax2.south west);
\draw [dashed] (c2) -- (ax2.north west);
\end{tikzpicture}

%% file: figures/hist.tikz
\usepgfplotslibrary{groupplots}

\begin{tikzpicture}

\definecolor{color0}{rgb}{0.12156862745098,0.466666666666667,0.705882352941177}

\begin{groupplot}[group style={group size=3 by 1, horizontal sep=1cm}]

\nextgroupplot[
tick align=outside,
tick pos=left,
x grid style={white!69.0196078431373!black},
xlabel={$\rho^v$ for $\mc C_1$},
xmajorgrids,
xmin=0, xmax=9,
xtick style={color=black},
y grid style={white!69.0196078431373!black},
ylabel={Number of vehicles},
ymajorgrids,
ymin=0, ymax=30,
ytick style={color=black},
height=5cm,
width=6cm,
area style,
]
\addplot+[color0,ybar interval,mark=no] plot coordinates { (0.67, 5.0) (0.94, 12) (1.21, 24) (1.49, 26) (1.76, 7) (2.03, 7) (2.31, 5) (2.59, 17) (2.86, 19) (3.13, 17) (3.41, 4) (3.68, 4) (3.96, 1) (4.23, 0) (4.51, 1) (4.78, 0) (5.06, 0) (5.33, 2) (5.60, 1) (5.88, 3) (6.15, 2) (6.43, 0) (6.71, 1) (6.98, 1) (7.26, 2) (7.53, 0) (7.81, 1) (8.08, 0) (8.35, 0) (8.63, 2)};

\nextgroupplot[
tick align=outside,
tick pos=left,
x grid style={white!69.0196078431373!black},
xlabel={$\rho^v$ for $\mc C_2$},
xmajorgrids,
xmin=0, xmax=8,
xtick style={color=black},
y grid style={white!69.0196078431373!black},
ymajorgrids,
ymin=0, ymax=30,
ytick style={color=black},
height=5cm,
width=6cm,
area style,
]
\addplot +[color0,ybar interval,mark=no] plot coordinates { (1.00, 12) (1.24, 10) (1.48, 14) (1.72, 20) (1.95, 16) (2.19, 5) (2.43, 12) (2.67, 13) (2.91, 13) (3.15, 5) (3.39, 13) (3.62, 6) (3.86, 1) (4.10, 1) (4.34, 1) (4.58, 0) (4.82, 0) (5.06, 0) (5.29, 0) (5.53, 2) (5.77, 2) (6.01, 0) (6.25, 0) (6.48, 1) (6.72, 1) (6.96, 0) (7.20, 0) (7.44, 0) (7.68, 1) (7.91, 3)};

\nextgroupplot[
tick align=outside,
tick pos=left,
x grid style={white!69.0196078431373!black},
xlabel={$\rho^v$ for $\mc C_3$},
xmajorgrids,
xmin=0, xmax=9,
xtick style={color=black},
y grid style={white!69.0196078431373!black},
ymajorgrids,
ymin=0, ymax=30,
ytick style={color=black},
height=5cm,
width=6cm,
area style,
]
\addplot+[color0,ybar interval,mark=no] plot coordinates {(1.47, 9) (1.70, 14) (1.93, 17) (2.16, 17) (2.39, 8) (2.61, 9) (2.84, 7) (3.08, 12) (3.3, 6) (3.53, 8) (3.76, 7) (3.99, 5) (4.22, 1) (4.45, 2) (4.67, 0) (4.91, 0) (5.13, 0) (5.36, 1) (5.59, 1) (5.82, 0) (6.05, 0) (6.28, 1) (6.51, 1) (6.73, 0) (6.96, 0) (7.19, 0) (7.42, 0) (7.65, 1) (7.88, 1) (8.11, 1)};

\end{groupplot}
\end{tikzpicture}